\documentclass[11pt,english]{article}
\usepackage{amsfonts}
\usepackage{float}
\usepackage{amsmath}
\usepackage{amssymb}
\usepackage{subcaption, graphicx}
\usepackage{setspace}
\usepackage[authoryear]{natbib}
\usepackage{appendix}
\usepackage{grffile}
\usepackage{natbib}
\usepackage{thmtools}
\usepackage{enumitem}

\newtheorem{proposition}{Proposition}
\newtheorem{example}{Example}
\newtheorem{remark}{Remark}
\allowdisplaybreaks
\input{tcilatex}
\begin{document}

\title{Nonparametric hierarchical Bayesian quantiles}
\author{\textsc{Luke Bornn} \\
\textit{Department of Statistics and Actuarial Science, Simon Fraser
University } \\
\texttt{bornn@fas.harvard.edu} \and \textsc{Neil Shephard} \\
\textit{Department of Economics and Department of Statistics, Harvard
University}\\
\texttt{shephard@fas.harvard.edu} \and \textsc{Reza Solgi} \\
\textit{Department of Statistics, Harvard University}\\
\texttt{rezasolgi@fas.harvard.edu} }
\maketitle

\begin{abstract}
Here we develop a method for performing nonparametric Bayesian inference on
quantiles. Relying on geometric measure theory and employing a Hausdorff
base measure, we are able to specify meaningful priors for the quantile
while treating the distribution of the data otherwise nonparametrically. We
further extend the method to a hierarchical model for quantiles of
subpopulations, linking subgroups together solely through their quantiles.
Our approach is computationally straightforward, allowing for censored and
noisy data. We demonstrate the proposed methodology on simulated data and an
applied problem from sports statistics, where it is observed to stabilize
and improve inference and prediction.
\end{abstract}

\renewcommand\thmcontinues[1]{Continued}

\renewcommand\thmcontinues[1]{Continued}


\baselineskip=20pt

\section{Introduction}

Consider learning about $\beta $, the $\tau \in (0,1)$ quantile of the
random variable $Z$. This will be based on data $\mathcal{D}%
=\{z_{1},...,z_{n}\}$, where we assume $z_{i}$, $i=1,2,...,n$, are scalars
and initially that they are independent and identically distributed. \ We
will  perform nonparametric Bayesian inference on $\beta $ given $\mathcal{D}
$. The importance of quantiles is emphasized by, for example, \cite%
{Parzen(79),Parzen(04)}, \cite{KoenkerBassett(78)} and \cite{Koenker(05)}.
By solving this problem we will also deliver a nonparametric Bayesian
hierarchical quantile model, which allows us to analyze data with
subpopulations only linked through quantiles. The methods extend to censored
and partially observed data. \ 

\subsection{Background}

In early work on Bayesian inference on quantiles, Section 4.4 of \cite%
{Jeffreys(61)} used a \textquotedblleft substitute
likelihood\textquotedblright\ $s(\beta )=\binom{n}{n_{\beta }}\tau
^{n_{\beta }}(1-\tau )^{n-n_{\beta }}$, where $n_{\beta
}=\sum_{i=1}^{N}1(z_{i}\leq \beta )$. See also \cite{BoosMonahan(86)}, \cite%
{Lavine(95)} and \cite{DunsonTaylor(05)}. This relates to other
approximations to the likelihood suggested by \cite{Lazar(03)}, \cite%
{LancasterJun(10)} and \cite{YangHe(12)}, who use empirical likelihoods, and 
\cite{ChernozhukovHong(03)} who are inspired by some connections with
M-estimators. \cite{ChamberlainImbens(03)} use a Bayesian bootstrap (\cite%
{Rubin(81)}) to carry out Bayesian inference on a quantile but have no
control over the prior for $\beta $.

\cite{YuMoyeed(01)} carried out Bayesian analysis of quantiles using a
likelihood based on an asymmetric Laplace distribution for the regression
residuals $e_{i}=y_{i}-x_{i}^{\prime }\beta $ (see also \cite%
{KoenkerMachado(99)} and \cite{Tsionas(03)}), $L(\mathcal{D}|\beta )=\exp
\{-\sum_{i=1}^{n}\rho _{\tau }(e_{i})\}$ 
where $\rho _{\tau }(\cdot )$ is the \textquotedblleft check
function\textquotedblright\ (\cite{KoenkerBassett(78)}), 
\begin{equation}
\rho _{\tau }(e)=|e|\left\{ (1-\tau )1_{e<0}+\tau 1_{e\geq 0}\right\} ,\quad
e\in R.  \label{check function}
\end{equation}%
Here $\rho _{\tau }(e)$ is continuous everywhere, convex and differentiable
at all points except when $e=0$. This Bayesian posterior is relatively easy
to compute using mixture representations of Laplace distributions. Papers
which extend this tradition include \cite{KozumiKobayashi(11)}, \cite%
{LiXiLin(10)}, \cite{Tsionas(03)}, \cite{KottasKrnjajic(09)} and \cite%
{YangWangHe(15)}. Unfortunately the Laplace distribution is a misspecified
distribution and so typically yields inference which is overly optimistic. 
\cite{YangWangHe(15)} and \cite{FengChenHe(15)} discuss how to overcome some
of these challenges, see the related works by \cite{ChernozhukovHong(03)}
and \cite{Muller(13)}.

Closer to our paper is \cite{HjortPetrone(07)} who assume the distribution
function of $Z$ is a Dirichlet process with parameter $aF_{0}$, focusing on
when $a\downarrow 0$. \cite{HjortWalker(09)} write down nonparametric
Bayesian priors on the quantile function. Our focus is on using informative
priors for $\beta $, but our focus on a non-informative prior for the
distribution of $Z$ aligns with that of \cite{HjortPetrone(07)}.

Our paper is related to \cite{BornnShephardSolgi(16)} who develop a Bayesian
nonparametric approach to moment based estimation. Their methods do not
cover our case; the differences are brought out in the next section. The
intellectual root is similar though: the quantile model only specifies a
part of the distribution, so we complete the model by using Bayesian
nonparametrics.

Hierarchical models date back to \cite{Stein(66)}, while linear regression
versions were developed by \cite{LindleySmith(72)}. Discussions of the
literature include \cite{MorrisLysy(12)} and \cite{Efron(10)}. Our focus is
on developing models where the quantiles of individual subpopulations are
thought of as drawn from a common population-wide mixing distribution, but
where all other features of the subpopulations are nonparametric and
uncommon across the populations. The mixing distribution is also
nonparametrically specified. There is some linkages with deconvolution
problems, see for example \cite{ButuceaComte(09)} and \cite%
{CavalierHengartner(09)}, but our work is discrete and not linear. It is
more related to, for example, \cite{Robbins(56)}, \cite{CarlinLouis(08)}, 
\cite{McAuliffeBleiJordan(06)} and \cite{Efron(13)} on empirical Bayes
methods.

Here we report a simple to use method for handling this problem, which
scales effectively with the sample size and the number of subpopulations.
The method extends to allow for censored data. Our hierarchical method is
illustrated on an example drawn from sports statistics.

\subsection{Outline of the paper}

In Section \ref{sect:one pop} we discuss our modelling framework and how we
define Bayesian inference on quantiles. Particular focus is on uniqueness
and priors. A flexible way of building tractable models is developed. This
gives an analytic expression for the posterior on a quantile. A Monte Carlo
analysis is carried out to study the bias, precision and coverage of our
proposed method, which also compares the results to that seen for sample
quantiles using central limit theories and bootstraps. In Section \ref%
{sect:hierarch} we extend the analysis by introducing a nonparametric
hierarchical quantile model and show how to handle it using very simple
simulation methods. A detailed study is made of individual sporting careers
using the hierarchical model, borrowing strengths across careers when the
careers are short and data is limited. In Section \ref{sect:truncate} we
extend the analysis to data which is censored and this is applied in
practice to our sporting career example. Section \ref{section:conclusions}
concludes, while an Appendix contains various proofs of results stated in
the main body of the paper. \ 

\section{A Bayesian nonparametric quantile\label{sect:one pop}}

\subsection{Definition of the problem}

We use the conventional modern definition of the $\tau $ quantile $\beta$,
that is 
\begin{equation*}
\beta =\underset{b}{\func{argmin}}\ \mathrm{E}\left\{ \rho _{\tau
}(Z-b)\right\} .
\end{equation*}%
To start suppose $Z$ has known finite support $\mathcal{S}%
=\{s_{1},...,s_{J}\}$, and write 
\begin{equation*}
\Pr (Z=s_{j}|\theta )=\theta _{j},\quad \text{for }1\leq j\leq J,
\end{equation*}%
with $\theta =(\theta _{1},\theta _{2},...,\theta _{J-1})^{\prime }\in
\Theta _{\theta }$, and $\Theta _{\theta }\subseteq \Delta$, where $\Delta$
is the simplex, $\Delta=\{\theta ; \ \iota ^{\prime }\theta <1 \text{ and }
\theta _{j}>0\}$, and define $\theta _{J}=1-\iota ^{\prime }\theta $, in
which $\iota $ is a vector of ones. The function%
\begin{equation*}
\Psi (b,\theta )=\mathrm{E}_{\theta }\left\{ \rho _{\tau }(Z-b)\right\}
=\sum_{j=1}^{J}\theta _{j}\rho _{\tau }(s_{j}-b),
\end{equation*}%
is continuous everywhere, convex and differentiable at all points except
when $b\in \mathcal{S}$.

We define the \textquotedblleft Bayesian nonparametric
quantile\textquotedblright\ problem as learning from data  the unknowns 
\begin{equation*}
(\beta ,\theta ^{\prime })^{\prime }\in \Theta _{\beta ,\theta },\quad \text{%
where}\quad \Theta _{\beta ,\theta }\subseteq \mathbb{R}\times \Delta
\subset \mathbb{R}^{J}.
\end{equation*}

Each point within $\Theta _{\beta ,\theta }$ is a pair $(\beta ,\theta )$
which satisfies both the probability axioms and 
\begin{equation*}
\beta =\underset{b}{\func{argmin}}\ \sum_{j=1}^{J}\theta _{j}\rho _{\tau
}(s_{j}-b).
\end{equation*}%
Here $\theta $ almost surely determines $\beta $ --- this will be formalized
in Proposition \ref{Prop:uniqueness of beta}.

\begin{figure}[!ht]
\centering
\begin{minipage}[c]{0.66\textwidth}
    \centering
    Unique $\beta$ \\
    (with probability $1$) \\ 
    \vspace{0.4cm}
    \includegraphics[height=10cm]{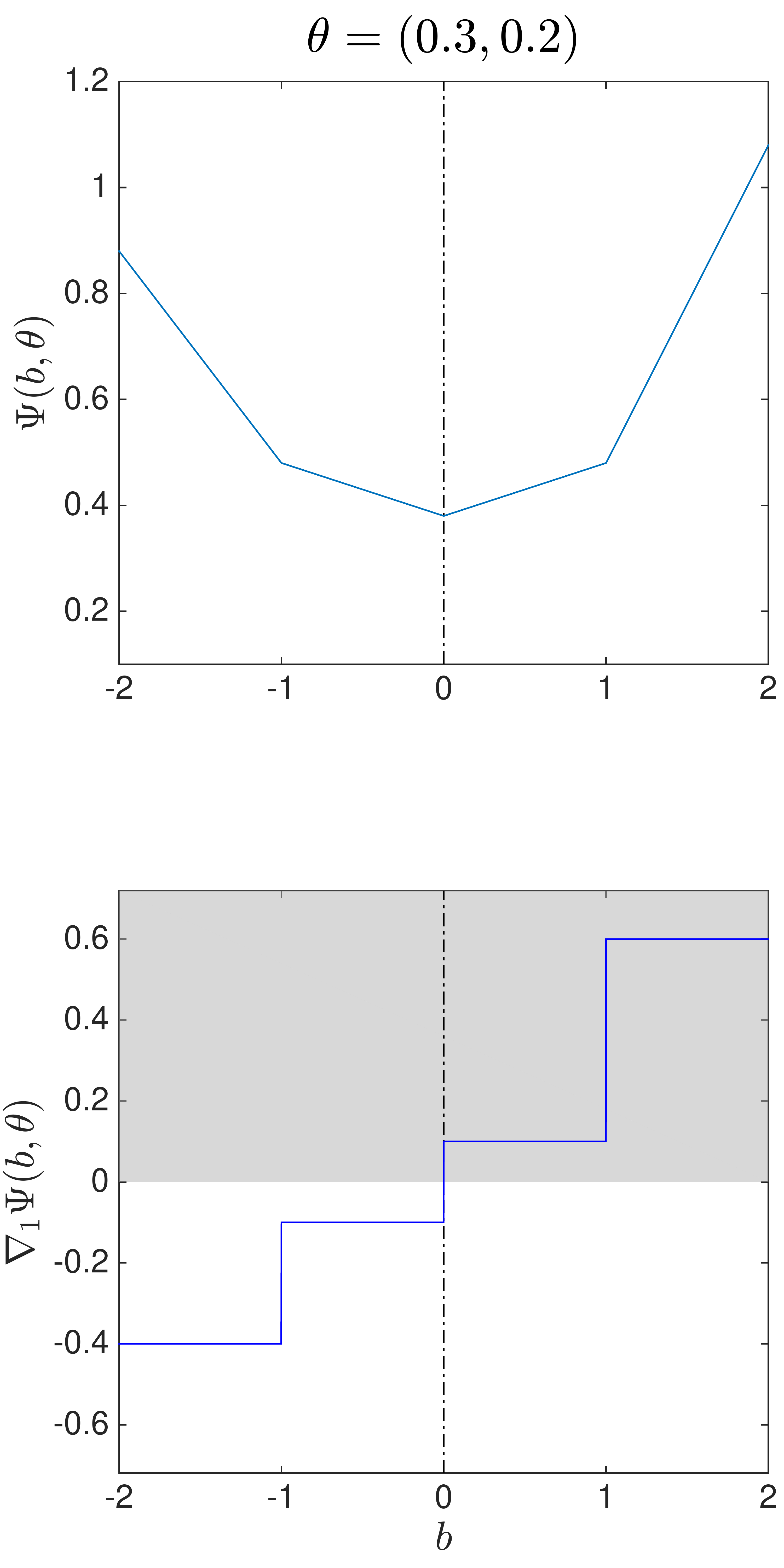}
    \includegraphics[height=10cm]{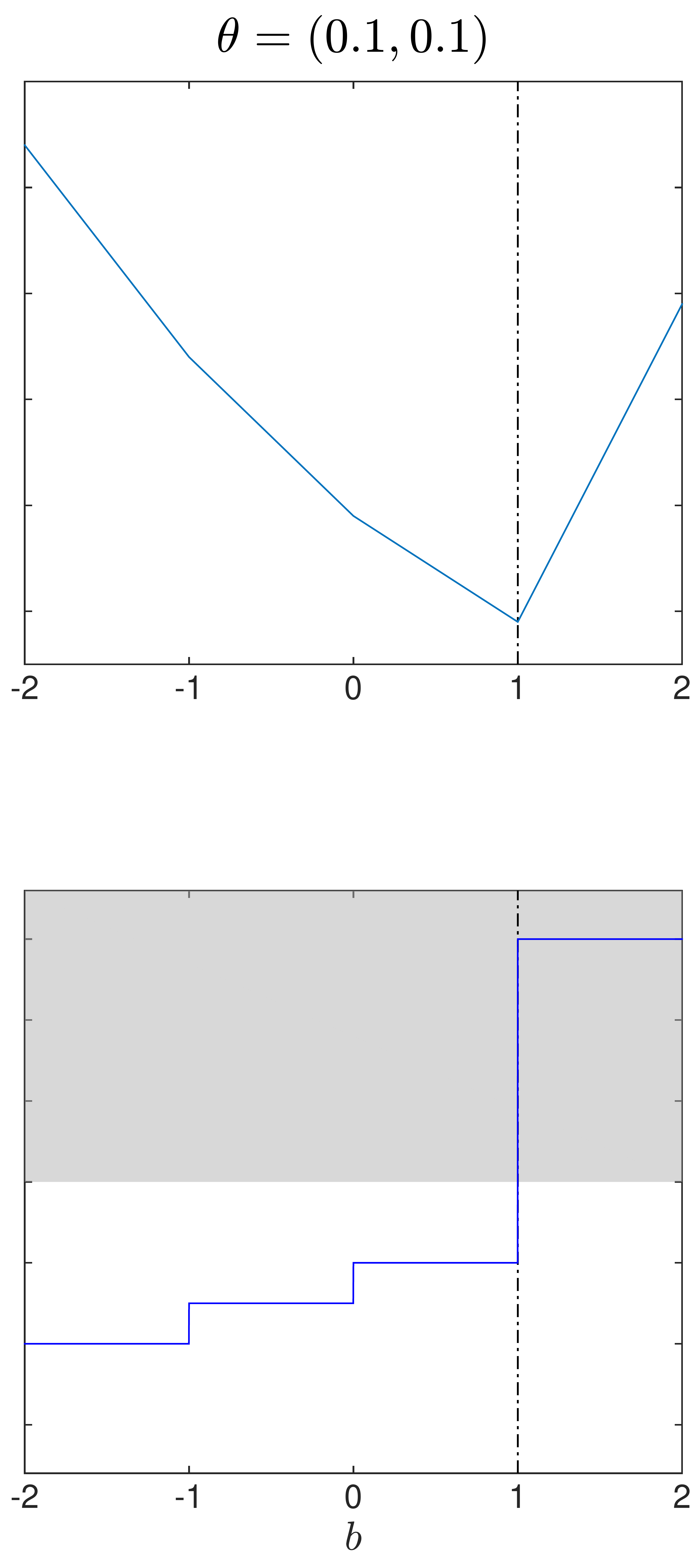}
\end{minipage}
\vline
\begin{minipage}[c]{0.33\textwidth}
    \centering
    Non-unique $\beta$ \\
    (with probability $0$) \\ 
    \vspace{0.4cm}
    \includegraphics[height=10cm]{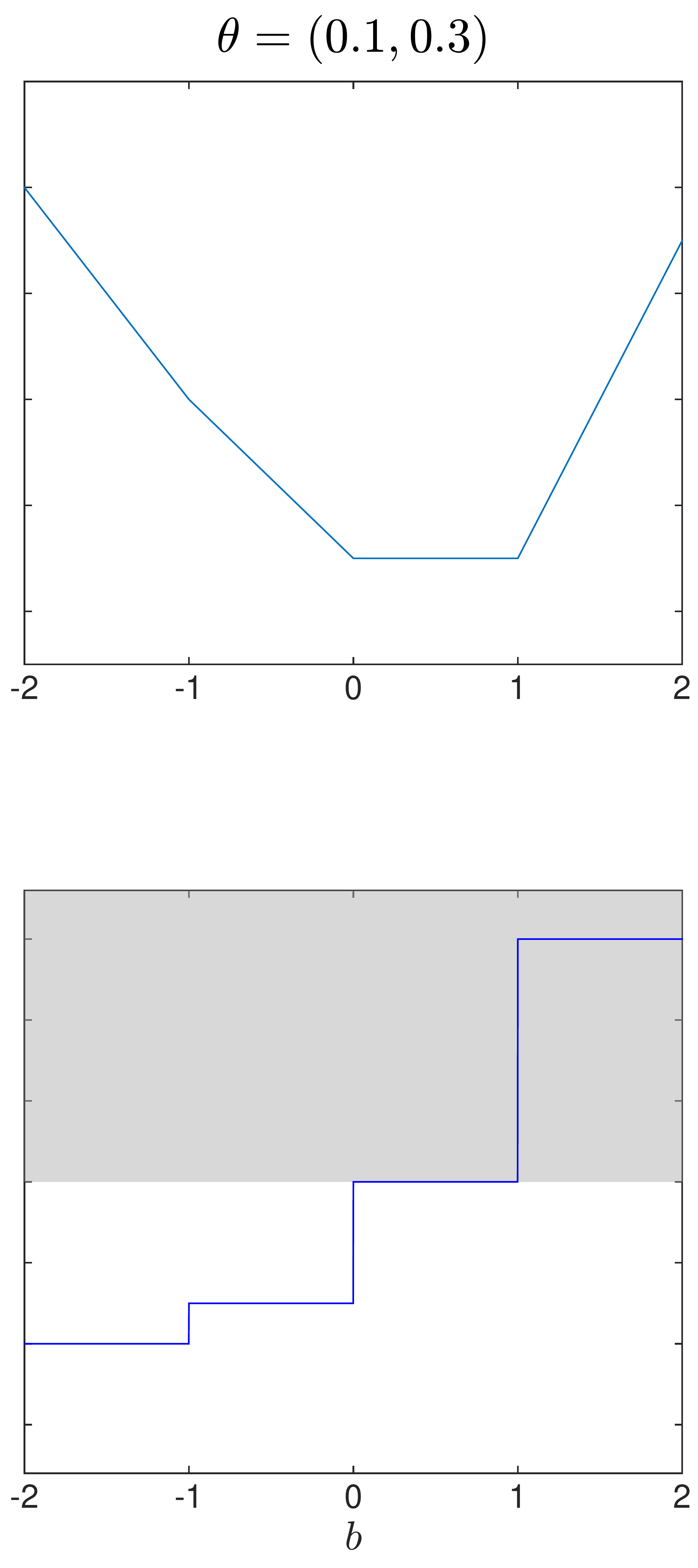}
\end{minipage}
\caption{{\protect\small This plot shows the 0.4-quantile with support }$%
S=\{-1,0,1\}${\protect\small . Plotted is }$\Psi (b,\protect\theta )$%
{\protect\small \ and its directional derivatives with respect to }$b$%
{\protect\small , }$\protect\nabla _{1}\Psi (b,\protect\theta )$%
{\protect\small . Left hand has }$\protect\theta =(0.3,0.2)^{\prime }$, the
center is $\protect\theta =(0.1,0.1)^{\prime }$, {\protect\small \ and the
right hand is }$\protect\theta =(0.1,0.3)^{\prime }${\protect\small . In the
left and center, the quantiles are are $0$ and $1$, respectively, while, in
the right the optimization does not have a unique solution. \ }}
\label{Fig:QE_Example1}
\end{figure}

\begin{example}
\label{exa:QEExample1} Figure \ref{Fig:QE_Example1}\footnote{%
Figure \ref{Fig:QE_Example1} demonstrates that Bayesian nonparametric
quantile estimation is not a special case of Bayesian nonparametric $\psi $
type M-estimators, and so not a special case of moment estimation. This
means we are outside the framework developed by \cite{BornnShephardSolgi(16)}%
.} sets $\tau =0.4$, and $\mathcal{S}=\{-1,0,1\}$. In the left panel, for $%
\theta =(0.3,0.2)^{\prime }$, we plot $\Psi (b,\theta )$ and its directional
derivatives\footnote{%
Recall, for the generic function $f(b)$, the corresponding directional
derivative is $\nabla _{v}f(b)=\lim_{h\downarrow 0}\frac{f(b+hv)-f(b)}{h}$.
\ } with respect to $b$, $\nabla _{1}\Psi (b,\theta )$. The resulting
quantile is $\beta =0$. In the center panels, $\theta =(0.1,0.1)^{\prime }$,
implying $\beta =1$. \ $\beta $ is not unique iff $\theta _{1}=0.4$ or $%
\theta _{1}+\theta _{2}=0.4$ --- which are 0 probability events. An example
of the latter case is $\theta =\left( 0.1,0.3\right) ^{\prime }$, which is
shown in the right panel . Here $\Psi (b,\theta )$ is minimized on $[0,1]$%
\footnote{%
If $\mathcal{D}=\mathcal{S}$ then the empirical quantile is $\widehat{\beta }%
=\underset{b}{\func{argmin}}\ \sum_{j=1}^{J}\rho _{\tau }(s_{j}-b)$, which
is non-unique if $\tau J$ is an integer (e.g. if $\tau =0.5$, then if $J$ is
even).}.
\end{example}

Proposition \ref{Prop:uniqueness of beta} formalizes the connection between $%
\beta $ and $\theta $. 

\begin{proposition}
\label{Prop:uniqueness of beta}\ Without loss of generality, assume $%
s_{1}<\cdots <s_{J}$. Then $\beta $ is unique iff $\tau \notin \{\theta
_{1},\theta _{1}+\theta _{2},....,\theta _{1}+\cdots +\theta _{J-1}\}$. If $%
\theta $ has a continuous distribution with respect to the Lebesgue measure,
then with probability 1, for each $\theta $ there is a unique quantile $%
\beta \in \mathcal{S}$ and with probability 1%
\begin{equation}
\frac{\partial \beta }{\partial \theta ^{\prime }}=0.\newline
\label{der beta against theta}
\end{equation}
\end{proposition}

Proposition \ref{Prop:uniqueness of beta} means we can partition the simplex
in $J+1$ sets, $\Delta=\left( \bigcup_{k=1}^{J}\mathcal{A}_{k}\right) \cup 
\mathcal{N}$, where $\mathcal{N}$ is a zero Lebesgue measure set and the
sets $\mathcal{A}_{k}=\{\theta \in \Delta;\ s_{k}=\underset{b}{\func{argmin}}%
\ \Psi (b,\theta )\}$, $1\leq k\leq J$, contain all the values of $\theta $
which deliver a quantile $\beta =s_{k}=\underset{b}{\func{argmin}}\ \Psi
(b,\theta )$. We write this compactly as $\beta =t(\theta )$, $\beta \in 
\mathcal{S}$, $\theta \in \Delta$, and the corresponding set index $%
k=k(\theta )$, $1\leq k\leq J$, $\theta \in \Delta$, so $\beta =s_{k(\theta
)}$.

\begin{figure}[tbp]
\begin{center}
\includegraphics[width=8cm]{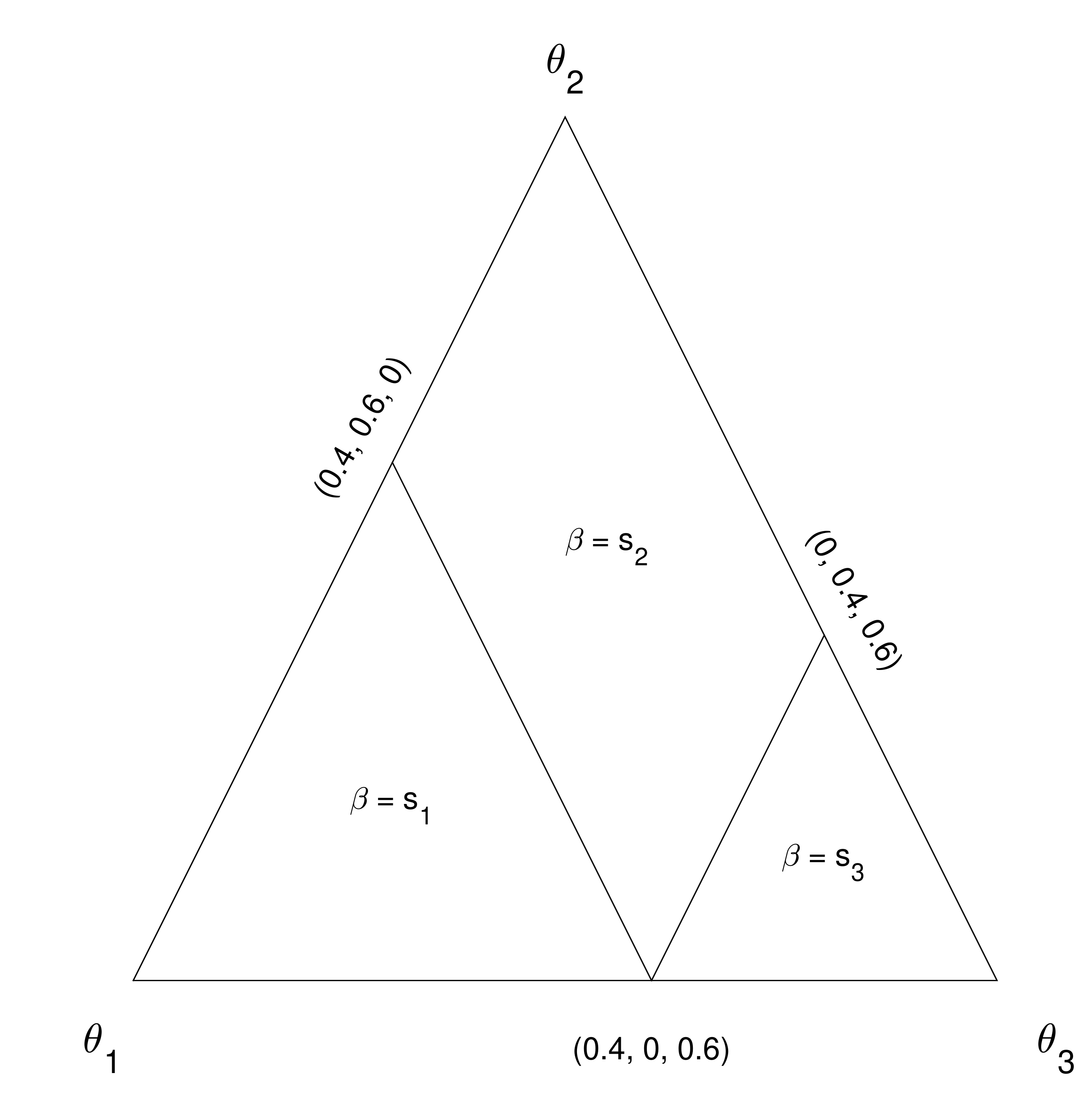}
\end{center}
\caption{Ternary plots of $\protect\theta _{1},\protect\theta _{2}$ and $%
\protect\theta _{3}=1-\protect\theta _{1}-\protect\theta _{2}$ and the
implied quantiles $\protect\beta $ at level $\protect\tau =0.4$. \ Here $%
\protect\beta \in \left\{ s_{1},s_{2},s_{3}\right\} $. $\mathcal{A}_{k}$ is
the set of probabilities $\protect\theta _{1},\protect\theta _{2}$ where $%
\protect\beta =s_{k}$. }
\label{Fig:TernaryPlots1}
\end{figure}

\begin{example}[continues=exa:QEExample1]
Figure \ref{Fig:TernaryPlots1} is a ternary plot showing all possible values
of $\theta =\left( \theta _{1},\theta _{2}\right) ^{\prime }$ and $\theta
_{3}=1-\theta _{1}-\theta _{2}$\ and the implied value of $\beta $ overlaid
for $\tau =0.4$. \ The values of $\theta $ which contain distinct values of $%
\beta $ are collected into the sets $\mathcal{A}_{1}$ (where $\beta =s_{1}$%
), $\mathcal{A}_{2}$ (where $\beta =s_{2}$), $\mathcal{A}_{3}$ (where $\beta
=s_{3}$). The interior lines marking the boundaries between these sets are
the zero measure events collected into $\mathcal{N}$. The union of the
disjoint sets $\mathcal{A}_{1},\mathcal{A}_{2},\mathcal{A}_{3}$, and $%
\mathcal{N}$, make up the simplex $\Delta$. \ 
\end{example}

\subsection{The prior and posterior}

The set of admissible pairs $(\beta ,\theta )$ is denoted by $\Theta _{\beta
,\theta }\subseteq \mathcal{S}\times \Delta$. Now $\Theta _{\beta ,\theta }$
is a lower dimensional space as $\beta =t(\theta )$. Using the Hausdorff
measure\footnote{%
Assume $E\subseteq \mathbb{R}^{n}$, $d\in \lbrack 0,+\infty )$ and $\delta
\in (0,+\infty ]$. The Hausdorff premeasure of $E$ is defined as follows, 
\begin{equation*}
\mathcal{H}_{\delta }^{d}(E)=v_{m}\ \inf_{\substack{ E\subseteq \cup E_{j} 
\\ d(E_{j})<\delta }}\ \ \sum_{j=1}^{\infty }\left( \frac{\text{diam}(E_{j})%
}{2}\right) ^{d}
\end{equation*}%
where $v_{m}=\frac{\Gamma (\frac{1}{2})^{d}}{2^{d}\Gamma (\frac{d}{2}+1)}$
is the volume of the unit $d$-sphere, and $\text{diam}(E_{j})$ is the
diameter of $E_{j}$. $\mathcal{H}_{\delta }^{d}(E)$ is a nonincreasing
function of $\delta $, and the $d$-dimensional Hausdorff measure of $E$ is
defined as its limit when $\delta $ goes to zero, $\mathcal{H}^{d}(E)=\lim 
_{\substack{ \delta \rightarrow 0^{+}}}\ \mathcal{H}_{\delta }^{d}(E)$. The
Hausdorff measure is an outer measure. Moreover $\mathcal{H}^{n}$ defined on 
$\mathbb{R}^{n}$ coincide with Lebesgue measure. See \cite{Federer(69)} for
more details. \ }, we are able to assign measures to the lower dimensional
subsets of $\mathcal{S}\times \Delta$, and therefore we can define
probability density functions with respect to Hausdorff measure on manifolds
within $\mathcal{S}\times \Delta$. \ 

One approach to building a joint prior $p(\beta ,\theta )$ is to place a
prior on $\beta \in \mathcal{S}$, which we write as 
\begin{equation*}
p(\beta ),\quad \beta \in \mathcal{S}
\end{equation*}%
and then build a conditional prior density, 
\begin{equation*}
p(\theta |\beta =s_{k}),\quad \theta \in \mathcal{A}_{k}
\end{equation*}%
recalling $\mathcal{A}_{k}\subseteq \Delta$. Then the joint density with
respect to Hausdorff measure on $\Theta _{\beta ,\theta }$ is 
\begin{equation*}
p(\beta ,\theta )=p(\beta )p(\theta |\beta ).
\end{equation*}

For the quantile problem, with probability one $\beta =t(\theta)$, so the
\textquotedblleft area formula" of \cite{Federer(69)} (see also \cite%
{DiaconisHolmesShahshahani(13)} and \cite{BornnShephardSolgi(16)}) implies
the marginal density for the probabilities is induced as 
\begin{equation*}
p(\theta )=p(\beta ,\theta ),\quad \beta =t(\theta ),
\end{equation*}%
as Proposition \ref{Prop:uniqueness of beta} shows that $\partial \beta
/\partial \theta ^{\prime }=0$. Here the right hand side is the density of
the prior with respect to Hausdorff measure defined on $\Theta _{\beta
,\theta }$, while the left hand side is the implied density of the prior
distribution of $\theta $ with respect to Lebesgue measure defined on the
simplex $\Delta$.

The model's likelihood is, 
\begin{equation*}
\prod_{j=1}^{J}\theta _{j}^{n_{j}},
\end{equation*}%
where $n_{j}=\sum_{i=1}^{n}1(z_{i}=s_{j})$. Then the posterior distribution
of $\beta ,\theta $ will be, 
\begin{equation}
p(\theta |\mathcal{D})=p(\beta =s_{k(\theta )},\theta |\mathcal{D})\propto
p(\beta =s_{k(\theta )})p(\theta |\beta =s_{k(\theta
)})\prod_{j=1}^{J}\theta _{j}^{n_{j}},\quad \beta =t(\theta ).
\label{posterior}
\end{equation}%
This means that 
\begin{equation*}
p(\beta =s_{k}|\mathcal{D})=\int_{\mathcal{A}_{k}}p(\theta |\mathcal{D})%
\mathrm{d}\theta \propto p(\beta =s_{k})\int_{\mathcal{A}_{k}}\left\{
p(\theta |\beta =s_{k})\prod_{j=1}^{J}\theta _{j}^{n_{j}}\right\} \mathrm{d}%
\theta .
\end{equation*}

\subsection{A class of $p(\protect\theta |\protect\beta )$ models}

Assume $f_{\Delta }(\theta )$ is the density function of a continuous
distribution on $\Delta $, and define, 
\begin{equation*}
c_{k}=\Pr_{f_{\Delta }}(\beta =s_{k})=\int_{\mathcal{A}_{k}}f_{\Delta
}(\theta )\mathrm{d}\theta .
\end{equation*}%
Then one way to build an explicit prior for $p(\theta |\beta )$ is to decide
to set 
\begin{equation*}
p(\theta |\beta =s_{k})=\frac{f_{\Delta }(\theta )}{c_{k}}1_{\mathcal{A}%
_{k}}(\theta ),\quad \theta \in \mathcal{A}_{k}.
\end{equation*}%
Proposition \ref{Prop Dirich c} shows how to compute $\left\{ c_{k}\right\} $%
.

\begin{proposition}
\label{Prop Dirich c}Here $c_{1}=1-\Pr (\theta _{1}<\tau )$, $c_{J}=\Pr
\left( \sum_{j=1} \theta _{j}<\tau \right) $, and $c_{k}=\Pr \left(
\sum_{j=1}^{k-1}\theta _{j}<\tau \right) -\Pr \left( \sum_{j=1}^{k}\theta
_{j}<\tau \right) $, for $k=2,...,J-1$.
\end{proposition}

This conditional distribution can be combined with a fully flexible prior $%
\Pr (\beta =s_{k})=b_{k}$, where $b_{k}>0$, for $1\leq k\leq J$, and $%
\sum_{k=1}^{J}b_{k}=1$. Returning to the general case, this implies the
joint 
\begin{equation}
p(\theta )=p(\beta =s_{k(\theta )},\theta )=\frac{b_{k(\theta )}}{%
c_{k(\theta )}}f_{\Delta }(\theta ),  \label{our marginal}
\end{equation}%
which in turn means, $\Pr (\beta =s_{k})=\int_{\mathcal{A}_{k}}p(\beta
,\theta ;\alpha )\mathrm{d}\theta =b_{k}$, the scientific marginal for $%
\beta $. Note that $p(\theta )$ is discontinuous at the set boundaries (that
is the zero Lebesgue measure set $\mathcal{N}$), and $p(\theta )\neq
f_{\Delta }(\theta )$ unless $b_{k}=c_{k}$ for all $k$.

From (\ref{posterior}) the posterior distribution of $\beta ,\theta $ will
be, 
\begin{equation*}
p(\beta =s_{k(\theta )},\theta |\mathcal{D})\propto \frac{b_{k(\theta )}}{%
c_{k(\theta )}}f_{\Delta }(\theta )\prod_{j=1}^{J}\theta _{j}^{n_{j}},\quad 
\text{and\quad }p(\beta =s_{k}|\mathcal{D})\propto \frac{b_{k}}{c_{k}}\int_{%
\mathcal{A}_{k}}\left\{ f_{\Delta }(\theta )\prod_{j=1}^{J}\theta
_{j}^{n_{j}}\right\} \mathrm{d}\theta .
\end{equation*}%
The Dirichlet case is particularly convenient. \ 

\subsection{Dirichlet special case}

Let $f_{\Delta }$ be the Dirichlet density, $f_{D}(\theta ;\alpha )=B(\alpha
)^{-1}\prod_{j=1}^{J}\theta _{j}^{\alpha _{j}-1}$, where $\alpha =(\alpha
_{1},...,\alpha _{J})$ is the vector of positive parameters, and $B(\alpha )$
is the beta function. Then $c_{k}$ can be computed via Proposition \ref{Prop
Dirich c} using the distribution function\footnote{%
So $\Pr \left( \theta _{k}^{+}<\tau \right) = I_{\tau }(\alpha
_{k}^{+},\alpha _{J}^{+}-\alpha _{k}^{+}) = B_k$, in which $I_{\tau}(\alpha,
\beta) = B(\tau ,\alpha, \beta)/B(\alpha, \beta)$ is the regularized
incomplete beta function, $B(\tau,\alpha,\beta) = \int_{0}^{\tau }x^{\alpha
-1}\left( 1-x\right) ^{\beta -1}\mathrm{d}x$ is the incomplete beta
function. 
When $\alpha _{k}^{+}$ and $\alpha _{J}^{+}-\alpha _{k}^{+}$ are large some
care has to be taken in computing $c_{k}$. We have written $%
c_{k}=B_{k-1}-B_{k}=B_{k}\left\{ \frac{B_{k-1}}{B_{k}}-1\right\}
=B_{k}\left\{ \exp \left( \log B_{k-1}-\log B_{k}\right) -1\right\} $ so $%
\log c_{k}=\log B_{k}+\log \left\{ \exp \left( \log B_{k-1}-\log
B_{k}\right) -1\right\} $. Now $B(x,a,b) =$ $_{2}F_{1}(a+b,1,a+1,x)\frac{1}{a%
}x^{a}(1-x)^{b}$ where $_{2}F_{1}$ is the Gauss hypergeometric function.
Hence we can compute $\log c_{k}$ accurately.} of 
\begin{equation*}
\theta _{k}^{+}\sim Be\left( \alpha _{k}^{+},\alpha _{J}^{+}-\alpha
_{k}^{+}\right) ,\quad \text{where generically\quad }\alpha
_{k}^{+}=\sum_{j=1}^{k}\alpha _{j}.
\end{equation*}

To mark their dependence on $\alpha $, in the Dirichlet case we write $%
c_{k}=c_{k}(\alpha )$. We will refer to \ 
\begin{equation}
p(\theta |\beta =s_{k})=\frac{f_{D}(\theta ;\alpha )}{c_{k}}1_{\mathcal{A}%
_{k}}(\theta ),\quad \theta \in \mathcal{A}_{k},  \label{truncated dirichlet}
\end{equation}%
as the density function of $D_{J}(\alpha ,k)$, the Dirichlet distribution
truncated to $\mathcal{A}_{k}$.

This result can be used to power the following simple prior to posterior
calculation. \ 

\begin{proposition}
\label{Prop of posterior for beta}When $f_{\Delta }(\theta )=f_{D}(\theta
;\alpha )$, then 
\begin{equation*}
\Pr (\beta =s_{k}|\mathcal{D})=\frac{1}{C(\alpha ,\mathbf{n})}\frac{%
c_{k}(\alpha +\mathbf{n})}{c_{k}(\alpha )}\Pr (\beta =s_{k}),
\end{equation*}%
where $\mathbf{n}=(n_{1},...,n_{J})$. Here $C(\alpha ,\mathbf{n})$ is the
normalizing constant, which is computed via enumeration, $C(\alpha ,\mathbf{n%
})=\sum_{k=1}^{J}\frac{c_{k}(\alpha +\mathbf{n})}{c_{k}(\alpha )}b_{k}$.
Further, 
\begin{equation*}
p(\theta |\mathcal{D})=\frac{1}{C(\alpha ,\mathbf{n})}\frac{\Pr (\beta
=s_{k(\theta )})}{c_{k(\theta )}(\alpha )}f_{D}(\theta ;\alpha +\mathbf{n}).
\end{equation*}
\end{proposition}

The Bayesian posterior mean or quantiles of the posterior can be computed by
enumeration unless $J$ is massive, in which case simulation can be used. 
\newline

\begin{figure}[!ht]
\begin{minipage}[l]{0.49\textwidth}
\begin{center}
    Discrete prior for median \\
    \vspace{0.1cm}
    \includegraphics[height=6cm]{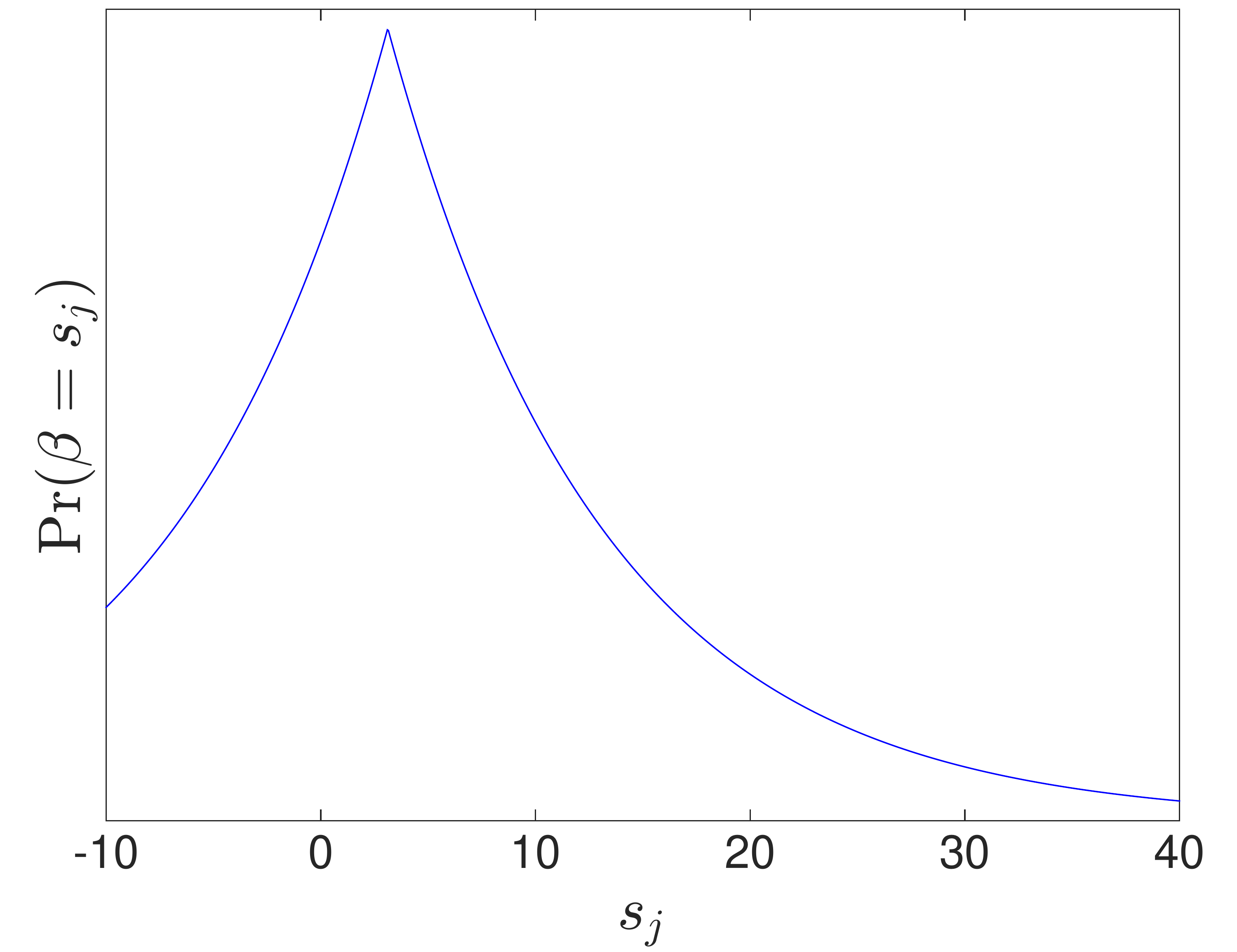}
\end{center}    
\end{minipage}
\begin{minipage}[r]{0.49\textwidth}
\begin{center}
    Posterior for median \\ 
    \vspace{0.1cm}
    \includegraphics[height=6cm]{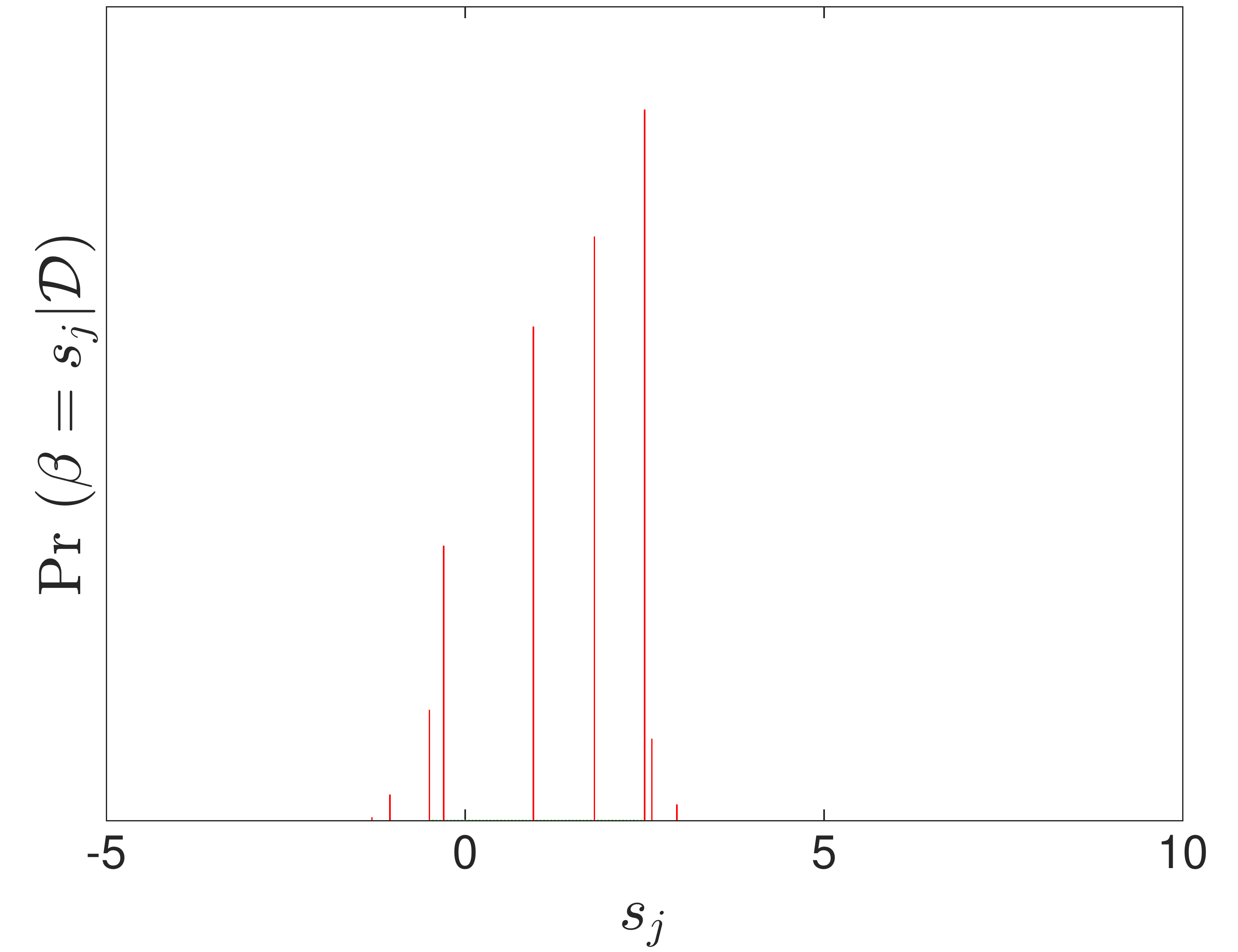}
\end{center}
\end{minipage}
\caption{{\protect\footnotesize {The left hand side shows the prior
distribution of $\protect\beta $ for the discrete method, and the right
shows the corresponding posterior for the first replication. Notice the
posterior has many small atoms marked in short green lines. These points
originate from the prior and represent around $1$ data points.}}}
\label{Fig:QE_4}
\end{figure}

\subsection{Monte Carlo experiment\ \label{sect: monte carlo experiment}\ \ }

Here $\mathcal{D}$ is simulated from the long right hand tailed $z_{i}%
\overset{iid}{\sim }-\log \chi _{1}^{2}$, so the $\tau $-quantile is $\beta
_{\tau }=-\log \left\{ F_{\chi _{1}^{2}}^{-1}(1-\tau )\right\} $. The
empirical quantile $\widehat{\beta }_{\tau }$ will be used to benchmark the
Bayesian procedures. The distribution of $\widehat{\beta }_{\tau }$ will be
computed using its limiting distribution $\sqrt{n}\left( \widehat{\beta }%
_{\tau }-\beta _{\tau }\right) \overset{d}{\rightarrow }N(0,\tau (1-\tau
)/f_{z}(\beta _{\tau })^{2})$ and by bootstrapping. In the case of the
limiting distribution, the $f_{z}(\beta _{\tau })$ has been estimated by a
kernel density estimator, with normal kernel and Silverman's optimal
bandwidth.

We build two Bayesian estimators:

\begin{enumerate}
\item \textbf{Discrete.} \ The $s_{j}=-10+50(j-1)/(J-1)$, where $J=1,000$,
and assume a weak prior for the median 
\begin{equation}
\Pr (\beta _{\tau }=s_{k})\propto \exp \left\{ -\lambda \left\vert
s_{k}-\delta _{\tau }\right\vert \right\} ,  \label{prior for beta}
\end{equation}%
where $\lambda =0.1$, and $\delta _{\tau }=\beta _{\tau }+\gamma _{\tau }$,
where $\gamma _{\tau }>0$, and we let $\gamma _{\tau }$ increases when $\tau 
$ deviates from $0.5$. This prior is not centered at the true value of the
quantile and is more contaminated for the tail quantiles. In particular in
our simulations we use $\gamma _{0.5}=2.333$ and $\gamma _{0.9}=6.032$. The
data is binned using the support, and $\alpha =\frac{1}{J}$. The (\ref{prior
for beta}), for $\tau =0.5$, is shown in Figure \ref{Fig:QE_4} together with
the associate posterior for one replication of simulated data.

\item \textbf{Data.} \ The support $\mathcal{S}$ is the data (therefore $J=n$%
), $\alpha = \frac1J$, and the prior height (\ref{prior for beta}) sits on
those $J$ points of support (so the prior changes in each replication).

\end{enumerate}

Table \ref{tab:quantileInf} reports \ 
\begin{table}[tbp]
\centering{\footnotesize \ 
\begin{tabular}{l||rrrr|rrrr}
& \multicolumn{4}{c}{$\tau =0.5$} & \multicolumn{4}{|c}{$\tau =0.9$} \\ 
\hline
& \multicolumn{2}{c}{Sample quantile} & \multicolumn{2}{c}{Posterior} & 
\multicolumn{2}{|c}{Sample quantile} & \multicolumn{2}{c}{Posterior} \\ 
\hline
& CLT & Boot & Discrete & Data & CLT & Boot & Discrete & Data \\ \hline\hline
& \multicolumn{8}{c}{$n=10$} \\ \hline
Bias & -0.157 & 0.152 & 0.345 & 0.206 & 1.245 & -0.174 & 1.083 & -0.167 \\ 
$n^{1/2}$ SE & 2.221 & 2.119 & 2.155 & 2.136 & 7.909 & 5.087 & 4.764 & 5.206
\\ 
RMSE & 0.720 & 0.687 & 0.764 & 0.706 & 2.794 & 1.618 & 1.856 & 1.655 \\ 
Coverage & 0.913 & 0.943 & 0.936 & 0.897 & 0.805 & 0.645 & 0.932 & 0.638 \\ 
\hline\hline
& \multicolumn{8}{c}{$n=40$} \\ \hline
Bias & -0.039 & 0.038 & 0.085 & 0.054 & -0.224 & 0.050 & 0.438 & 0.098 \\ 
$n^{1/2}$ SE & 2.309 & 2.184 & 2.214 & 2.203 & 5.639 & 5.403 & 5.731 & 5.560
\\ 
RMSE & 0.367 & 0.347 & 0.360 & 0.353 & 0.919 & 0.856 & 1.007 & 0.885 \\ 
Coverage & 0.945 & 0.945 & 0.937 & 0.940 & 0.810 & 0.912 & 0.944 & 0.910 \\ 
\hline\hline
& \multicolumn{8}{c}{$n=160$} \\ \hline
Bias & 0.020 & 0.010 & 0.021 & 0.014 & 0.072 & 0.014 & 0.102 & 0.025 \\ 
$n^{1/2}$ SE & 2.358 & 2.258 & 2.266 & 2.262 & 6.130 & 5.650 & 5.767 & 5.700
\\ 
RMSE & 0.187 & 0.179 & 0.180 & 0.179 & 0.490 & 0.447 & 0.467 & 0.451 \\ 
Coverage & 0.955 & 0.948 & 0.946 & 0.953 & 0.922 & 0.944 & 0.952 & 0.947 \\ 
\hline\hline
& \multicolumn{8}{c}{$n=320$} \\ \hline
Bias & -0.003 & 0.002 & 0.005 & 0.003 & -0.015 & 0.002 & 0.023 & 0.004 \\ 
$n^{1/2}$ SE & 2.343 & 2.296 & 2.297 & 2.297 & 6.029 & 5.856 & 5.885 & 5.867
\\ 
RMSE & 0.093 & 0.091 & 0.091 & 0.091 & 0.239 & 0.231 & 0.234 & 0.232 \\ 
Coverage & 0.952 & 0.950 & 0.948 & 0.947 & 0.930 & 0.947 & 0.940 & 0.951%
\end{tabular}
}
\caption{{\protect\footnotesize {Monte Carlo experiment using 25,000
replications and a highly biased prior. Coverage probability is based on a
nominal 95\% confidence or credible interval. Bayesian estimators are the
posterior mean. RMSE denotes root mean square error. Boot denotes bootstrap,
CLT implemented using a kernel for the asymptotic standard error.}}}
\label{tab:quantileInf}
\end{table}
the results from $25,000$ replications, comparing the five different modes
of inference. It shows the asymptotic distribution of the empirical quantile
provides a poor guide when working within a thin tail even when the $n$ is
quite large. In the center of the distribution it is satisfactory by the
time $n$ hits 40. The bootstrap performs poorly in the tail when $n$ is
tiny, but is solid when $n$ is large. \ 

Not surprisingly the bootstrap of the empirical quantile $\widehat{\beta }$
and the Bayesian method using support from the data are very similar.
Assuming no ties, straightforward computations leads to, 
\begin{equation*}
\Pr (\widehat{\beta }=s_{j})=F_{\text{B}}\left( \lceil \tau J\rceil
-1;J,(j-1)/J\right) -F_{\text{B}}\left( \lceil \tau J\rceil -1;J,j/J\right)
\end{equation*}%
where $F_{\text{B}}(\cdot ;n,p)$ is the binomial cumulative distribution
function with size parameter $n$, and probability of success $p$.
Interestingly, for large $J$, this is a close approximation to $c_{j}(%
\mathbf{1})$. This connection will become more explicit in the next
subsection. \ 

The discrete Bayesian procedure is by far the most reliable, performing
quite well for all $n$. It does have a large bias for small $n$, caused by
the poor prior, but the coverage is encouraging. \ Overall, there is some
evidence that for small samples the Bayesian estimators perform well in
moderate to large samples. The two Bayesian procedures have roughly the same
properties.

\subsection{Comparison with Jeffrey's substitution likelihood}

Some interesting connections can be established by thinking of $\alpha $ as
being small. \ 

\begin{proposition}
\label{prop small alpha}Conditioning on the data, if $\alpha _{k}\downarrow
0 $, and $\frac{\alpha_k}{\alpha_l} \rightarrow 1$, then $c_{k}(\alpha )
\rightarrow \frac{1}{J}$, and, 
\begin{eqnarray*}
c_{k}(\alpha +\mathbf{n}) &\rightarrow &\sum_{j=n_{k-1}^{+}}^{n_{k}^{+}-1}f_{%
\text{B}}(j;n-1,\tau ),\quad k=1,2,...,J,
\end{eqnarray*}%
where, for $k=0,1,...,n$, $f_{\text{B}}(k;n,p)=$ $\binom{n}{k}%
p^{k}(1-p)^{n-k}$, is the binomial probability mass function with the size
parameter $n$, and the probability of success $p$, and $n_{0}^{+}=0$.
\end{proposition}

The reason why $n-1$, not $n$,appears in the limit of $c_{k}(\alpha +\mathbf{%
n})$ is that $\mathcal{S}$ only has $n$ elements so $j$ runs from $0$ to $n-1
$. The proposition means that if there are no ties in the data and $\mathcal{%
D}=\mathcal{S}$, then 
\begin{equation*}
\begin{array}{lll}
c_{k}(\alpha +\mathbf{n}) & \rightarrow & f_{\text{\textit{B}}}(k-1;J-1,\tau
),\quad k=1,2,...,J, \\ 
\Pr (\beta =s_{k}|\mathcal{D}) & \rightarrow & C(\mathbf{n}%
)^{-1}f(k-1;J-1,\tau )\Pr (\beta =s_{k}).%
\end{array}%
\quad
\end{equation*}%
Here $C(\mathbf{n})$ is the normalizing constant, computed via enumeration, $%
C(\mathbf{n})=\sum_{k=1}^{J}f(k-1;J-1,\tau )b_{k}$.

The result in Proposition \ref{prop small alpha} is close to, but different
from, Jeffrey's substitution likelihood $s(\beta )=f(k;J,\tau )$, for $%
s_{k}\leq \beta <s_{k+1}$ where $s_{0}=-\infty $ and $s_{J+1}=\infty $
(Jeffrey has $n+1$ categories to choose from, not $n$, as he allows data
outside the supposed $\mathcal{S}$). $s(\beta )$ is a piecewise constant,
non-integrable function (which means it needs proper priors to make sense)
in $\beta \in R$, while for us $\beta \in \mathcal{S}$ (and the posterior is
always proper). \ 

\subsection{Comparison with Bayesian bootstrap}

The prior and posterior distribution of $\beta $ in the Bayesian bootstrap
are $\Pr (\beta =s_{k})=c_{k}(\alpha )$ and $\Pr (\beta =s_{k}|\mathcal{D}%
)=c_{k}(\alpha +\mathbf{n})$, respectively. Therefore, Proposition \ref{Prop
of posterior for beta} demonstrates that the choice of $b_{k}=c_{k}(\alpha )$
delivers the Bayesian bootstrap (here the results are computed analytically
rather than via simulation). If a Bayesian bootstrap was run, each draw
would be weighed by $w_{k}=b_{k}/c_{k}(\alpha )$ to produce a Bayesian
analysis using a proper prior; $w_{k}$ is the ratio of the priors and does
not depend upon the data. Finally, Proposition \ref{prop small alpha}
implies that as $\alpha \downarrow 0$, so $c_{k}(\alpha )\rightarrow J^{-1}$%
. This demonstrates that, in the Bayesian bootstrap, the implied prior of $%
\beta $ is the uniform discrete distribution on the support of the data. In
many applications this is an inappropriate prior.

\begin{remark}
\label{remark:sim theta given data}To simulate from 
\begin{equation*}
p(\theta |\mathcal{D})=\frac{1}{C(\alpha ,\mathbf{n})}\frac{b_{k(\theta )}}{%
c_{k(\theta )}(\alpha )}f_{D}(\theta ;\alpha +\mathbf{n}),\quad b_{k}=\Pr
(\beta =s_{k}),
\end{equation*}%
write $m_{k}=\frac{b_{k}}{c_{k}}/C(\alpha ,\mathbf{n})$, $m_{k}^{\prime
}=b_{k}/c_{k}$, $M=\max \left( m_{1},...,m_{J}\right) $ and $M^{\prime
}=\max \left( m_{1}^{\prime },...,m_{J}^{\prime }\right) $. Now $p(\theta |%
\mathcal{D})\leq Mf_{D}(\theta ;\alpha +\mathbf{n})$, for any $\theta $. We
can sample from $p(\theta |\mathcal{D})$ by drawing from $\text{Dirichlet}%
(\alpha +\mathbf{n})$ and accepting with probability $m_{k(\theta
)}/M=m_{k(\theta )}^{\prime }/M^{\prime }$. The overall acceptance rate is $%
1/M$. If the prior on $\beta $ is weakly informative then $m_{k}^{\prime
}\simeq 1$ for each $k$, and so the acceptance rate $m_{k(\theta )}^{\prime
}/M^{\prime }\simeq 1$.\ 
\end{remark}

\subsection{A cheap approximation}

If $J$ is large, $\alpha \downarrow 0$ and no ties, then a central limit
theory for binomial random variables implies \ 
\begin{equation*}
\frac{1}{J}\log f_{\text{B}}(k-1;J-1,\tau )\simeq -\frac{\left( \frac{k-1}{%
J-1}-\tau \right) ^{2}}{2\tau \left( 1-\tau \right) },
\end{equation*}%
which should be a good approximation unless $\tau $ is in the tails, or $J$
is small. So the resulting trivial approximations to the main posterior
quantities are 
\begin{eqnarray*}
\widehat{\mathrm{E}}\left( \beta |\mathcal{D}\right)
&=&\sum_{j=1}^{J}w_{j}^{\ast }s_{j},\quad w_{k}^{\ast
}=w_{k}b_{k}/\sum_{j=1}^{J}w_{j}b_{j},\quad w_{k}=\exp \left\{ -\frac{%
J\left( \frac{k-1}{J-1}-\tau \right) ^{2}}{2\tau \left( 1-\tau \right) }%
\right\} , \\
\widehat{\mathrm{Var}}\left( \beta |\mathcal{D}\right)
&=&\sum_{j=1}^{J}w_{j}^{\ast }\left\{ s_{j}-\widehat{\mathrm{E}}\left( \beta
|\mathcal{D}\right) \right\} ^{2},\quad \widehat{F}_{\beta |\mathcal{D}%
}(\beta )=\sum_{j=1}^{J}w_{j}^{\ast }1_{s_{j}\leq \beta }\text{.}
\end{eqnarray*}

When the prior is flat, this is a kernel weighted average of the data where
the weights are determined by the ordering of the data. So large weights are
placed on data with ranks $\left( k-1\right) /\left( J-1\right) $ which are
close to $\tau $. This is very close to the literature on kernel quantiles,
e.g. \cite{Parzen(79)}, \cite{Azzalini(81)}, \cite{Yang(85)} and \cite%
{SheatherMarron(90)}.

\section{Hierarchical quantile models\label{sect:hierarch}}

\subsection{Model structure}

Assume a population is indexed by $i=1,2,...,I$ subpopulations, and that our
random variable $Z$ again has known discrete support, $\mathcal{S}%
=\{s_{1},...,s_{J}\}$. \ Then we assume within the $i$-th subpopulation 
\begin{equation}
\Pr (Z=s_{j}|\theta ,i)=\theta _{j}^{(i)},  \label{cond independent}
\end{equation}%
thus allowing the distribution to change across the subpopulations. Here $%
\theta ^{(i)}=(\theta _{1}^{(i)},...,\theta _{J-1}^{(i)})$, $\theta
_{J}^{(i)}=1- \iota^{\prime (i)}$, and $\theta =(\theta ^{(1)},...,\theta
^{(I)})$. We assume the data $\mathcal{D}=\left\{ Z_{1},...,Z_{n}\right\} $
are conditionally independent draws from (\ref{cond independent}). We assume
that each time we see datapoints we also see which subpopulation the
datapoint comes from. The data from the $i$-th population will be written as 
$\mathcal{D}_{i}$.

For the $i$-th subpopulation, the Bayesian nonparametric $\tau $ quantile is
defined as 
\begin{equation*}
\beta _{i}=\underset{b}{\func{argmin}}\ \sum_{j=1}^{J}\theta _{j}^{(i)}\rho
_{\tau }(s_{j}-b).
\end{equation*}%
Collecting terms $\beta =\left( \beta _{1},\beta _{2},...,\beta _{I}\right)
^{\prime }$, the crucial assumption in our model is that 
\begin{equation*}
f(\theta|\beta )=\dprod\limits_{i=1}^{I}f(\theta ^{(i)}|\beta _{i}).
\end{equation*}%
This says the distributions across subpopulations are conditionally
independent given the quantile. That is, the single quantiles are the only
feature which is shared across subpopulations. \ 

We assume $\beta _{i}\in \mathcal{S}$, and the $\left\{ \beta _{i}\right\} $
are i.i.d. across $i$, but from the shared distribution $\Pr (\beta
_{i}=s_{j}|i,\pi )=\pi _{j}$, $i=1,2,...,I$, where $\pi =(\pi _{1},...,\pi
_{J-1})$, and $\pi _{J}=1- \iota^{\prime }\pi$. We write a prior on $\pi $
as $p(\pi )$. Then the prior on the hierarchical parameters is%
\begin{equation*}
f(\beta ,\pi )=f(\pi )f(\beta |\pi )=f(\pi )\dprod\limits_{i=1}^{I}f(\beta
_{i}|\pi ).
\end{equation*}%
This structured distribution will allow us to pool quantile information
across subpopulations.

Our task is to make inference on $\left( \beta _{1},\beta _{2},...,\beta
_{I}\right) ^{\prime }$ from $\mathcal{D}$. When taken together, we call
this a \textquotedblleft nonparametric hierarchical quantile
model\textquotedblright . This can also be thought of as related to the \cite%
{Robbins(56)} empirical Bayes method, but here each step is nonparametric.

By Bayes theorem, 
\begin{equation}
f(\beta ,\pi |\mathcal{D})\propto f(\beta ,\pi )f(\mathcal{D}|\beta ,\pi ).
\label{joint density}
\end{equation}%
We will access this joint density using simulation.

\begin{itemize}
\item \textbf{Algorithm 1: }$\beta ,\pi |\mathcal{D}$\textbf{\ Gibbs
sampler\ }
\end{itemize}

\begin{enumerate}
\item Sample from $\Pr (\beta |\mathcal{D},\pi )=\dprod\limits_{i=1}^{I}\Pr
(\beta _{i}|\mathcal{D}_{i},\pi )$.

\item Sample from $f(\pi |\mathcal{D},\beta )=f(\pi |\beta )$.
\end{enumerate}

In the Dirichlet case, we can sample from $\Pr (\beta _{i}|\mathcal{D}%
_{i},\pi )$ using Proposition \ref{Prop of posterior for beta}. \ If $f(\pi
) $ is Dirichlet, then $\pi |\beta =$Dirichlet$(\lambda +\nu )$, where $\nu
=(\nu _{1},...,\nu _{J})$, in which $\nu _{j}=\sum_{i=1}^{I}1(\beta
^{(i)}=s_{j})$. \ 

\subsection{Example: batting records in cricket\label{sect:cricket}}

We illustrate the hierarchical model using a dataset of the number of runs
(which is a non-negative integer) scored in each innings by the most recent
(by debut) $I=300$ English test players. \ \textquotedblleft
Tests\textquotedblright\ are international matches, typically played over 5
days. Here we look at only games involving the English national team. This
team plays matches against Australia, Bangladesh, India, New Zealand,
Pakistan, South Africa, Sri Lanka, West Indies and Zimbabwe. Batsmen can bat
up to twice in each test, but some players fail to get to bat in an
individual game due to the weather or due to the match situation. Some
players are elite batsmen and score many runs, others specialize in other
aspects of the game and have poor batting records without any runs. \ \ 

The database starts on 14th December 1951 and ends on 22nd January 2016.
Some of these players never bat, others have long careers, the largest of
which we see in our database is 235 innings, covering well over 100 test
matches. In test matches batsmen can continue their innings for potentially
a very long time and so can accumulate very high scores. An inning can be
left incomplete for a number of reasons, so the score is right-censored ---
such innings are marked as being \textquotedblleft not out\textquotedblright
. By the rules of cricket at least 9\% of the data must be right-censored.
The database is quite large, but has a simple structure. The statistical
challenge is with the data. Batting records are full of heterogeneity,
highly skewed, partially censored and heavy tailed data. It is a good test
case for our methods. \ 

Interesting academic papers on the statistics of batting includes \cite%
{KimberHansford(93)}, which is a sustained statistical analysis of
estimating the average performance of batsmen just using their own scores. 
\cite{Elderton(45)} is a pioneering cricket statistics paper in the same
spirit. More recent papers include \cite{PhilipsonBoys(15)} and \cite%
{Brewer(13)}.

Our initial aim will be to make inference on the $0.5$ quantile for each and
every batsmen, even if they have never batted. To start we will ignore the
\textquotedblleft not out\textquotedblright\ indicator. The player-by-player
empirical median ranges from $0$ and $46$, and is itself heavily negatively
skewed. \newline

\begin{figure}[!ht]
\begin{minipage}[l]{0.50\textwidth}
\begin{center}
    \hspace{0.75cm} $50$\% quantile \\
    \vspace{0.1cm}
    \includegraphics[height=6cm]{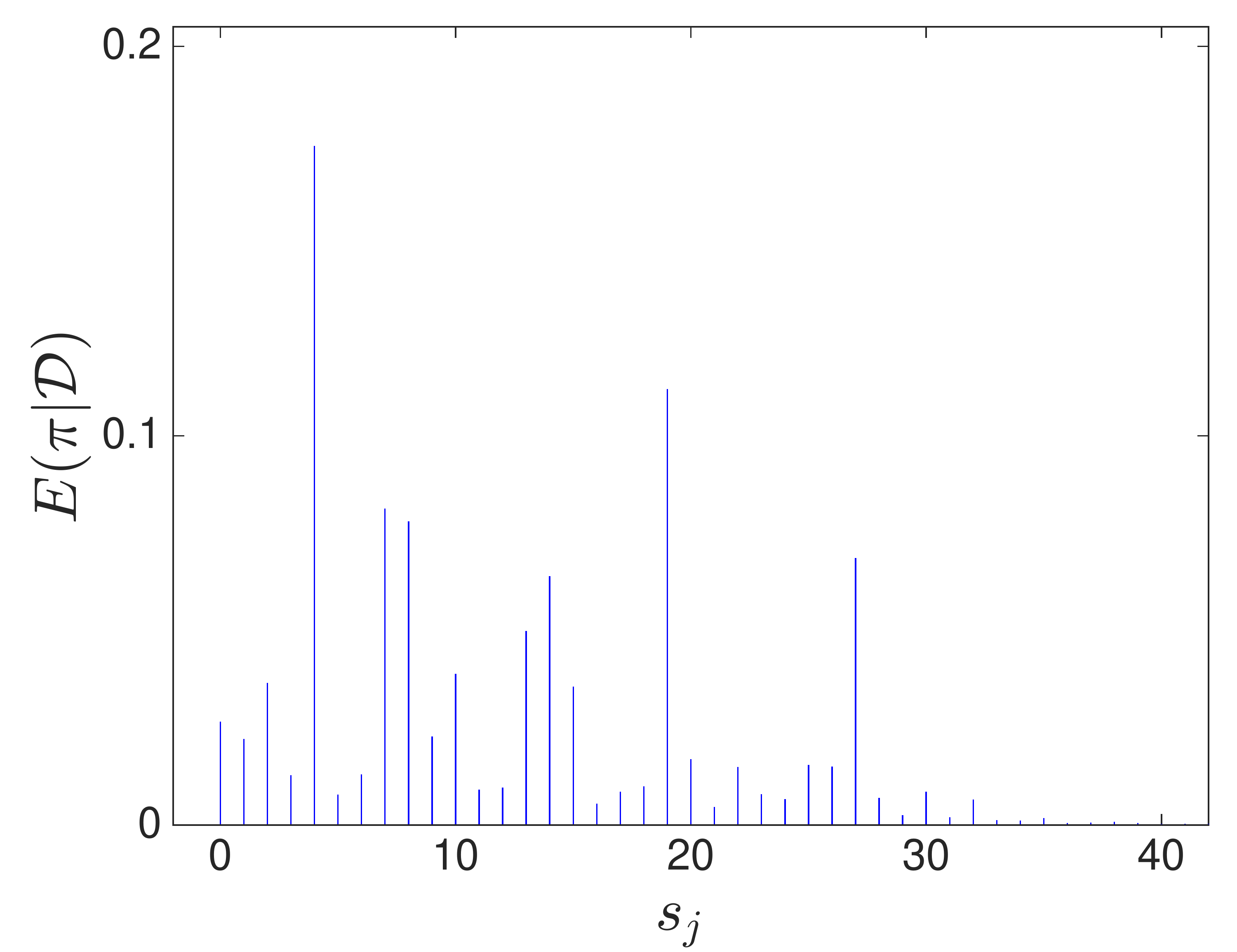}
\end{center}    
\end{minipage}
\begin{minipage}[r]{0.50\textwidth}
\begin{center}
    \hspace{0.85cm} Posterior on median for $4$ players
    \vspace{0.1cm}
    \includegraphics[height=6cm]{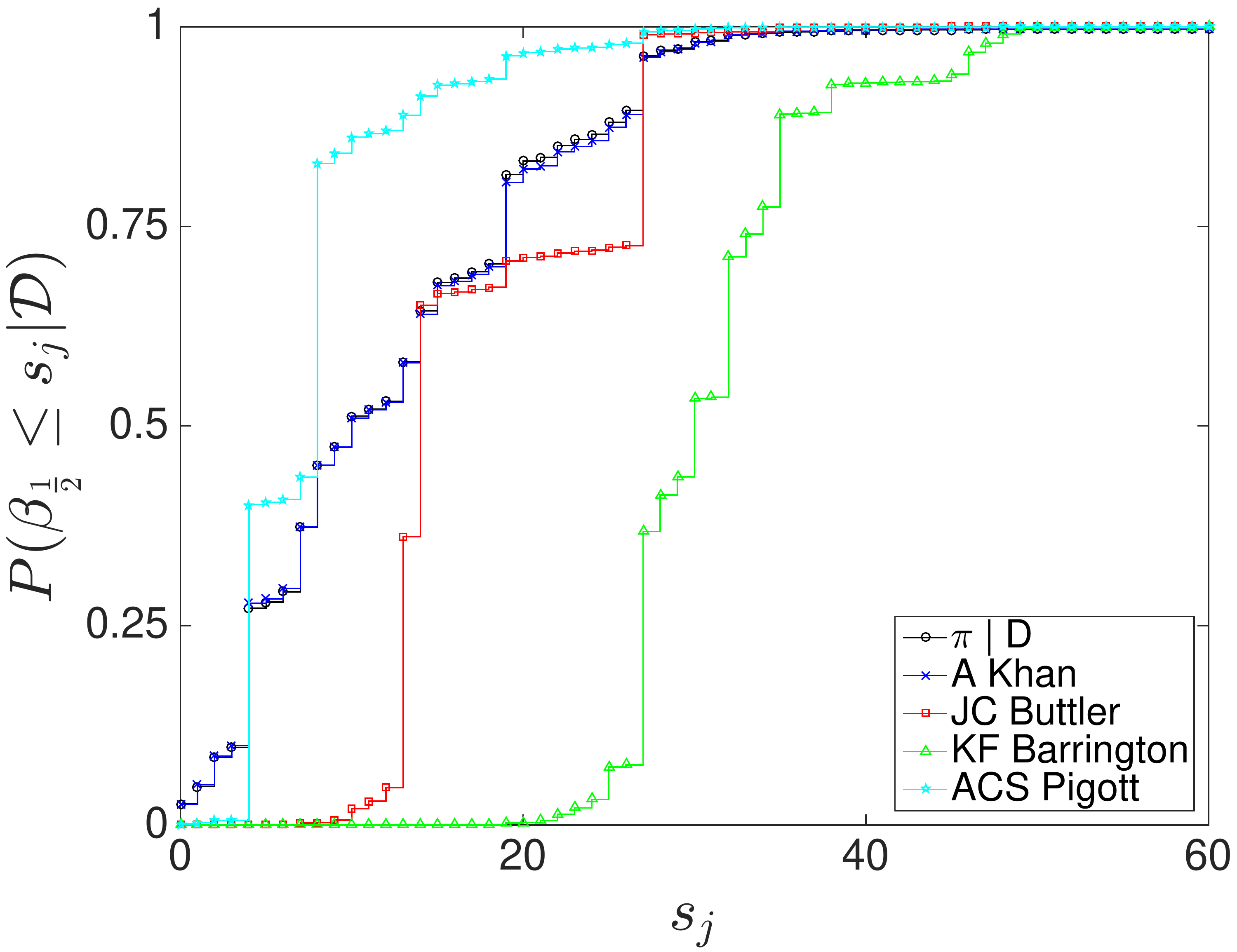}
\end{center}
\end{minipage}
\caption{{\protect\footnotesize The left hand side shows the posterior
distribution of the population probabilities of the quantiles }$\protect\pi %
_{j}=P(\protect\beta =s_{j})${\protect\footnotesize , }$\protect\tau =0.5$%
{\protect\footnotesize . The right hand side shows the posterior
distribution of median of several players along with the posterior
distribution of }$\protect\pi ${\protect\footnotesize . Notice in the case
of Barrington there is only one innings which finished in the range 36 to 44
inclusive, which makes estimating the median unexpectedly hard (given how
large a sample we have) and encourages the Bayesian method to aggressively
shrink the estimator of the median. \ } \ }
\label{fig:mixing distribution}
\end{figure}

The common support of data for all the players is $\mathcal{S}%
=\{0,1,...,350\}$, therefore $J=351$. The prior distribution of $\theta
^{(i)}$ is a Dirichlet distribution with $\alpha =(\alpha _{1},...,\alpha
_{J})$, where $\alpha _{j}=4\tilde{\alpha}_{j}+\frac{1}{J}$ with $\tilde{%
\alpha}_{j}\propto e^{-0.03s_{j}}$ and $\sum_{j=1}^{J}\tilde{\alpha}_{j}=1$
(The empirical probability mass function of batting scores of all English
players in the matches started between $1930$ and $1949$, $p_{i}=\Pr
(Z=s_{j})$, is approximately proportional to $e^{-0.03s_{j}}$. Therefore our
Dirichlet prior for $\theta $ is approximately centered around this
empirical probability mass function with a large variability). We assume $%
\pi \sim \text{Dirichlet}(\lambda )$, where $\lambda _{j}=\tilde{\lambda}%
_{j}+\frac{1}{J}$, in which $\tilde{\lambda}_{j}\propto e^{-\frac{1}{2}%
\left( \frac{s_{j}-15}{15}\right) ^{2}}$ for $j=1,...,J$, and $\sum_{j=1}^{J}%
\tilde{\lambda}_{j}=5$. In the left hand side of Figure \ref{fig:mixing
distribution} we have depicted $\mathrm{E}(\pi |\mathcal{D})$ for the $\tau
=0.5$ median case. Figure \ref{fig:two values of tau} shows the results for
the $\tau =0.3$ and $\tau =0.9$ cases. We will return to the non-median
cases in the next subsection. \ \ \ \ \ 
\begin{table}[!ht]
\centering{\footnotesize \ 
\begin{tabular}{l||rrr|rr|l||rrr|rr}
& \multicolumn{3}{c|}{Posterior} & \multicolumn{2}{c|}{} &  & 
\multicolumn{3}{c|}{Posterior} & \multicolumn{2}{c}{} \\ 
Batsman & $\widetilde{\beta }_{1/2}$ & Q$_{5}$ & Q$_{95}$ & $\widehat{\beta }%
_{1/2}$ & $n_{i}$ & Batsman & $\widetilde{\beta }_{1/2}$ & Q$_{5}$ & Q$_{95}$
& $\widehat{\beta }_{1/2}$ & $n_{i}$ \\ \hline
A Khan & 12.8 & 1 & 27 & -- & 0 & CJ Tavare & 17.4 & 13 & 25 & 19.5 & 56 \\ 
ACS Pigott & 7.8 & 4 & 19 & 6 & 2 & PCR Tufnell & 1.2 & 0 & 2 & 1 & 59 \\ 
A McGrath & 13.1 & 4 & 27 & 34 & 5 & MS Panesar & 1.6 & 0 & 4 & 1 & 64 \\ 
AJ Hollioake & 4.9 & 2 & 12 & 3 & 6 & CM Old & 8.4 & 7 & 11 & 9 & 66 \\ 
JB Mortimore & 16.7 & 9 & 19 & 11.5 & 12 & JA Snow & 5.5 & 4 & 8 & 6 & 71 \\ 
DS Steele & 18.8 & 7 & 38 & 43 & 16 & DW Randall & 14.6 & 9 & 19 & 15 & 79
\\ 
PJW Allott & 6.6 & 4 & 14 & 6.5 & 18 & RC Russell & 14.6 & 10 & 20 & 15 & 86
\\ 
JC Buttler & 17.4 & 13 & 27 & 13.5 & 20 & MR Ramprakash & 18.7 & 14 & 21 & 19
& 92 \\ 
W Larkins & 12.8 & 7 & 25 & 11 & 25 & PD Collingwood & 23.2 & 19 & 28 & 25 & 
115 \\ 
NG Cowans & 4.1 & 3 & 7 & 3 & 29 & RGD Willis & 4.1 & 4 & 5 & 5 & 128 \\ 
JK Lever & 5.4 & 4 & 10 & 6 & 31 & KF Barrington & 31 & 25 & 46 & 46 & 131
\\ 
M Hendrick & 3.4 & 1 & 4 & 2 & 35 & APE Knott & 17.6 & 13 & 24 & 19 & 149 \\ 
DR Pringle & 7.8 & 4 & 9 & 8 & 50 & IT Botham & 20.7 & 15 & 27 & 21 & 161 \\ 
C White & 11.5 & 7 & 19 & 10.5 & 50 & DI Gower & 26.9 & 25 & 28 & 27 & 204
\\ 
GO Jones & 15.9 & 10 & 22 & 14 & 53 & AJ Stewart & 25.6 & 19 & 28 & 27 & 235%
\end{tabular}
}
\caption{{\protect\footnotesize {\ Estimated median batting scores, treating
not outs as if they were completed innings (i.e. ignoring right censoring).
The batsman are ordered by sample size (i.e. the number of innings the
batsman had). Table shows, for each batsmen, the mean of the Bayesian
posterior of the median given the data, }}$\protect\widetilde{\protect\beta }%
_{1/2}$ = $E(\protect\beta _{1/2}|\mathcal{D})$, {\protect\footnotesize {the
sample median }}$\protect\widehat{\protect\beta }_{1/2}$ 
{\protect\footnotesize {and the sample size $n_{i}$. $Q_{5}$ and $Q_{95}$
are the estimates of the Bayesian 5\% and 95\% quantiles of the posterior
distribution of the median, so indicates how uncertain we are about the
Bayesian estimator of the mean. All the Bayesian quantities are estimated by
simulation. }}}
\label{tab:cricketSubpop}
\end{table}

In the right hand side plot in Figure \ref{fig:mixing distribution}, the
posterior distribution function of the median of scores for several players
have been compared with the posterior distribution function of $\pi $ (the
black curve). For the first player, A. Khan (the blue curve), no data are
available as he never batted, and the distribution is indistinguishable from
that for $\mathrm{E}(\pi |\mathcal{D)}$. A.C.S. Pigott played two innings
for England, scoring 4 and 8 not out. The light blue curve shows that even
with just two data points a lot of the posterior mass on the median has
moved to the left, but the median is very imprecisely estimated (the
estimate of median is $7.8$ with $95\%$ credible region $[4, 19]$). The red
curve corresponds to J. C. Buttler, whose sample median ($13$) is close to $%
\mathbb{E}_{\pi |\mathcal{D}}(\beta ) = 12.9$. His 20 actual scores were 85,
70, 45, 0, 59*, 13, 3*, 35*, 67, 14, 10, 73, 27, 7, 13, 11, 9, 12, 1, 42.
His scores are not particularly heavy-tailed and so the median is reasonably
well determined (the estimate of median is $17.4$ with $95\%$ credible
region $[13, 27]$). The green line shows the results for K. F. Barrington
who batted 131 times and one of the highest averages of any English batsman.
His median is relatively high ($31.0$) but surprisingly not well determined
(with $95\%$ credible region $[25, 46]$). Remarkably he has only once scored
between 36 and 44 (inclusive), so there is a whole range of possible scores
where there is no data. This stretches the Bayesian nonparametric interval.
The right hand side of Figure \ref{fig:mixing distribution} shows this
clearly. Of course a 90\% interval would be much shorter as it would not
include this blank range.\ \ \ 

Table \ref{tab:cricketSubpop} shows estimated posterior mean of the median
for 30 players, together with sample sizes, 90\% intervals, putting 5\% of
the posterior probability in each tail. Also given is the empirical median.
The players are sorted by sample size. It shows that when the sample size is
small there is a great deal of borrowing across the subpopulations. However,
when the subpopulation is large then the hierarchy does not make much
difference. McGrath's scores are 69, 81, 34, 4, 13 (with sample median $34$%
), so he has very little data in the middle (he either fails or scores
highly), and therefore the procedure shrinks the median a great deal towards
a typical median result (the Bayesian estimate is $13.1$). Steele's sample
median ($43$) is very high (it is very similar to Barrington's) and the
sample size is low ($16$). The resulting Bayes estimate is still a high
number ($18.8$), but is less than half of his sample median. Hence we think
the evidence is that Steele was a very good batsmen, but there is not the
evidence to rank him as a great batsman like Barrington. His record is more
in line with Botham and Ramprakash. \ \ 

\begin{figure}[!ht]
\begin{center}
\includegraphics[width=16cm]{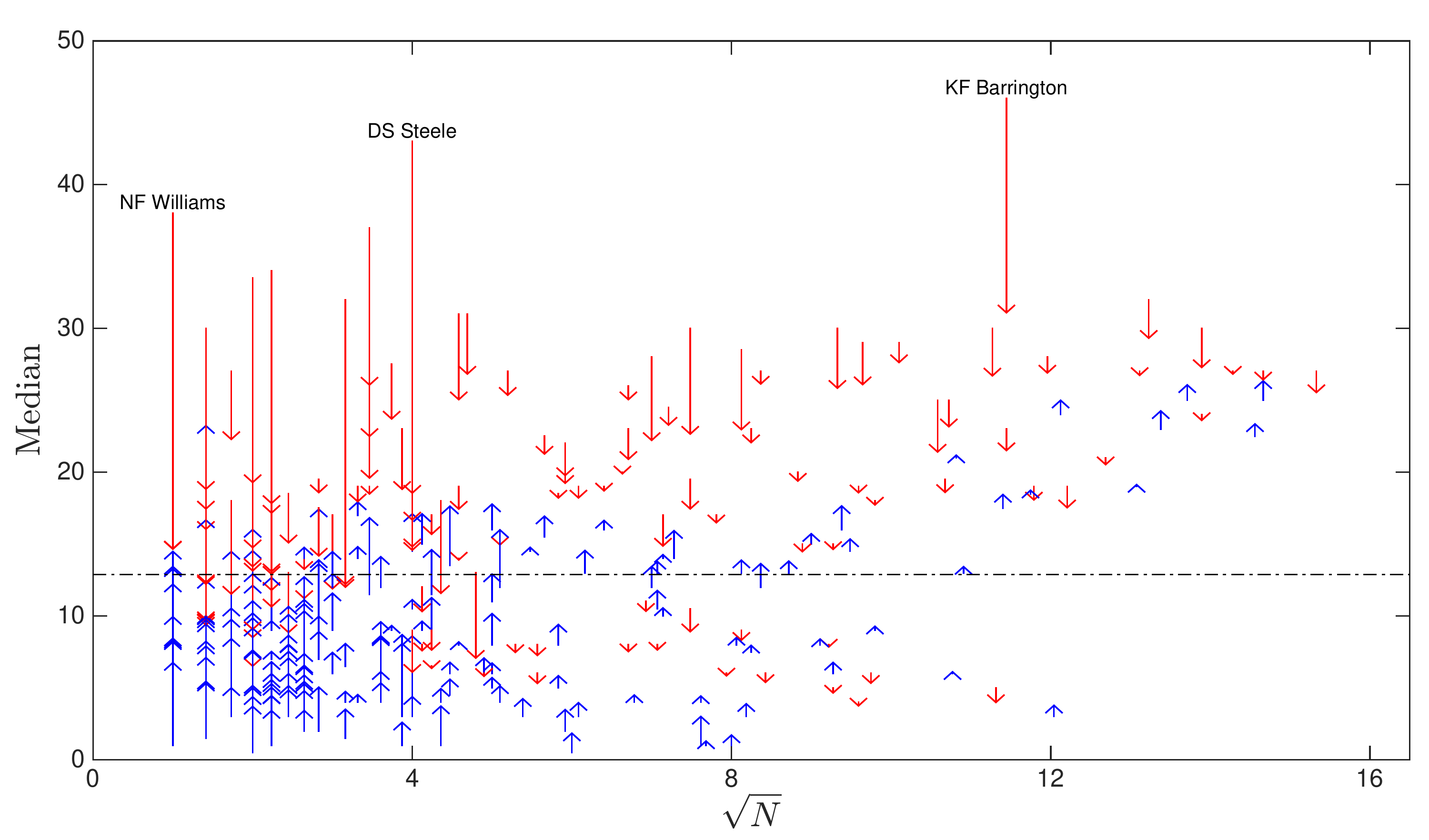}
\end{center}
\caption{{\protect\footnotesize Sample median (arrow nocks) and mean of
posterior distribution of medians (arrow heads) against the sample size for
all players. The blue arrows indicate the estimates which were moved
upwards, and the estimators which were moved down demonstrated by the red
arrows. The dashed line is the expected value of }$\protect\beta $%
{\protect\footnotesize \ under }$E(\protect\pi |D)${\protect\footnotesize .}}
\label{Fig:QE_Cricket45}
\end{figure}

Figure \ref{Fig:QE_Cricket45} highlights the shrinkage of the sample median
by the hierarchical model. We plot the batsman's sample median $\widehat{%
\beta }_{1/2}$ against the batsman's sample size $n_{i}$. Blue arrows show
that the Bayesian posterior mean of the median is below the sample median,
that is, it is shrunk down. Red arrows are the opposite, the Bayesian
estimator is above the sample median, so is moved upwards. The picture shows
there is typically more shrinkage for small sample sizes. But also, high
sample medians are typically shrunk more than low sample medians, but there
are more medians which are moved up than down. All this makes sense: the
data are highly skewed, so high scores can occur due to high medians or by
chance. Hence until we have seen a lot of high scores, we should shrink a
high median down towards a more common value. \ \ 

\subsection{Estimating the quantile function}

Of interest is $\beta _{\tau }$, the $\tau $-th quantile, as a function of $%
\tau $. Here we estimate that relationship pointwise, building a separate
hierarchical model for each value of $\tau $. The only change we will employ
is to set $\tilde{\lambda}_{j}\propto \exp \left\{ -\left( s_{j}-\mu _{\tau
}\right) ^{2}/\sigma _{\tau }^{2}\right\} $, allowing $\mu _{\tau
}=15+15\Phi ^{-1}(\tau )$ and $\sigma _{\tau }=15$. \newline

\begin{figure}[!ht]
\begin{minipage}[l]{0.49\textwidth}
\begin{center}
	\hspace{0.75cm} 30\% quantile \\
    \vspace{0.1cm}
    \includegraphics[height=6cm]{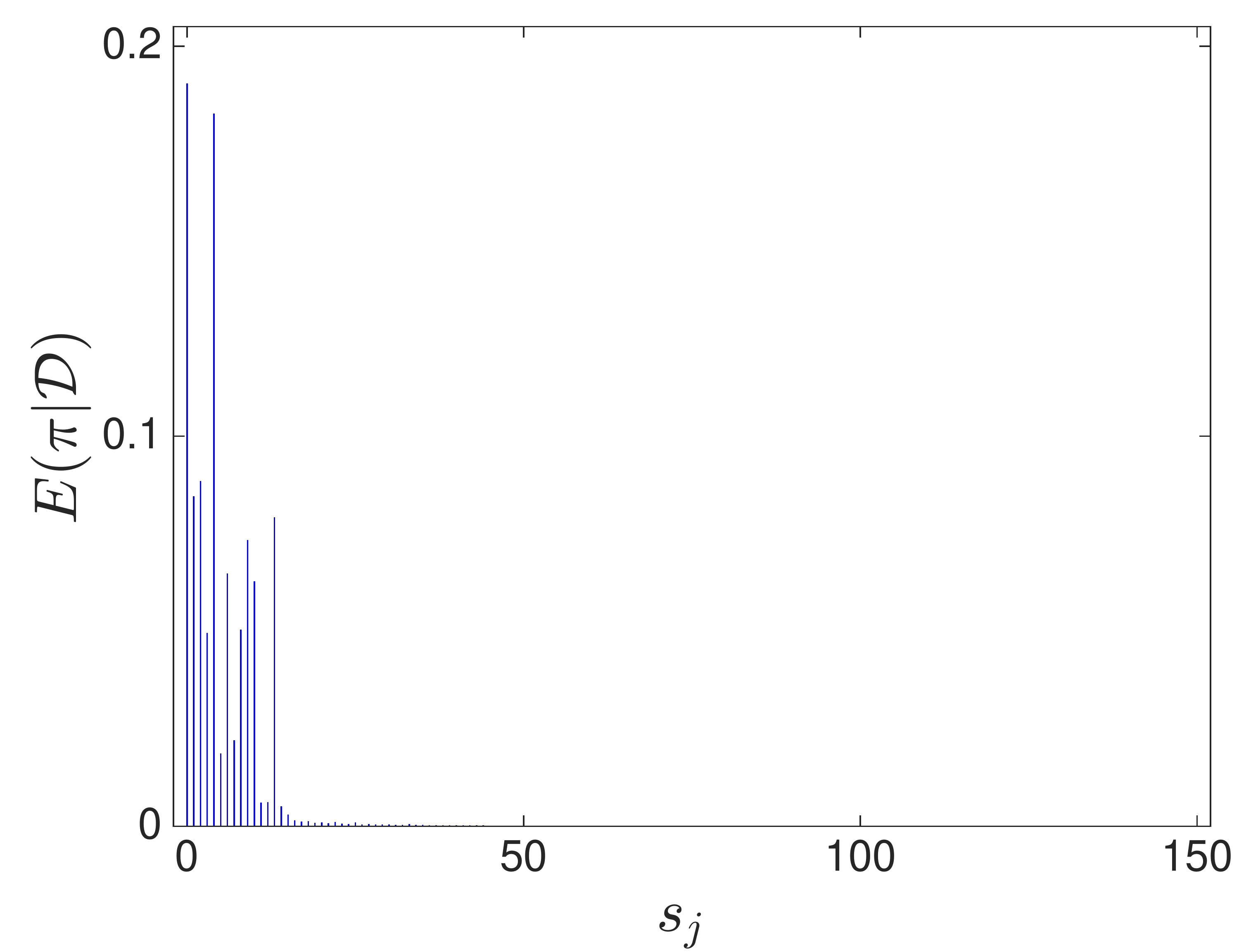}
\end{center}    
\end{minipage}
\begin{minipage}[r]{0.49\textwidth}
\begin{center}
    \hspace{0.75cm} 90\% quantile \\
    \vspace{0.1cm}
    \includegraphics[height=6cm]{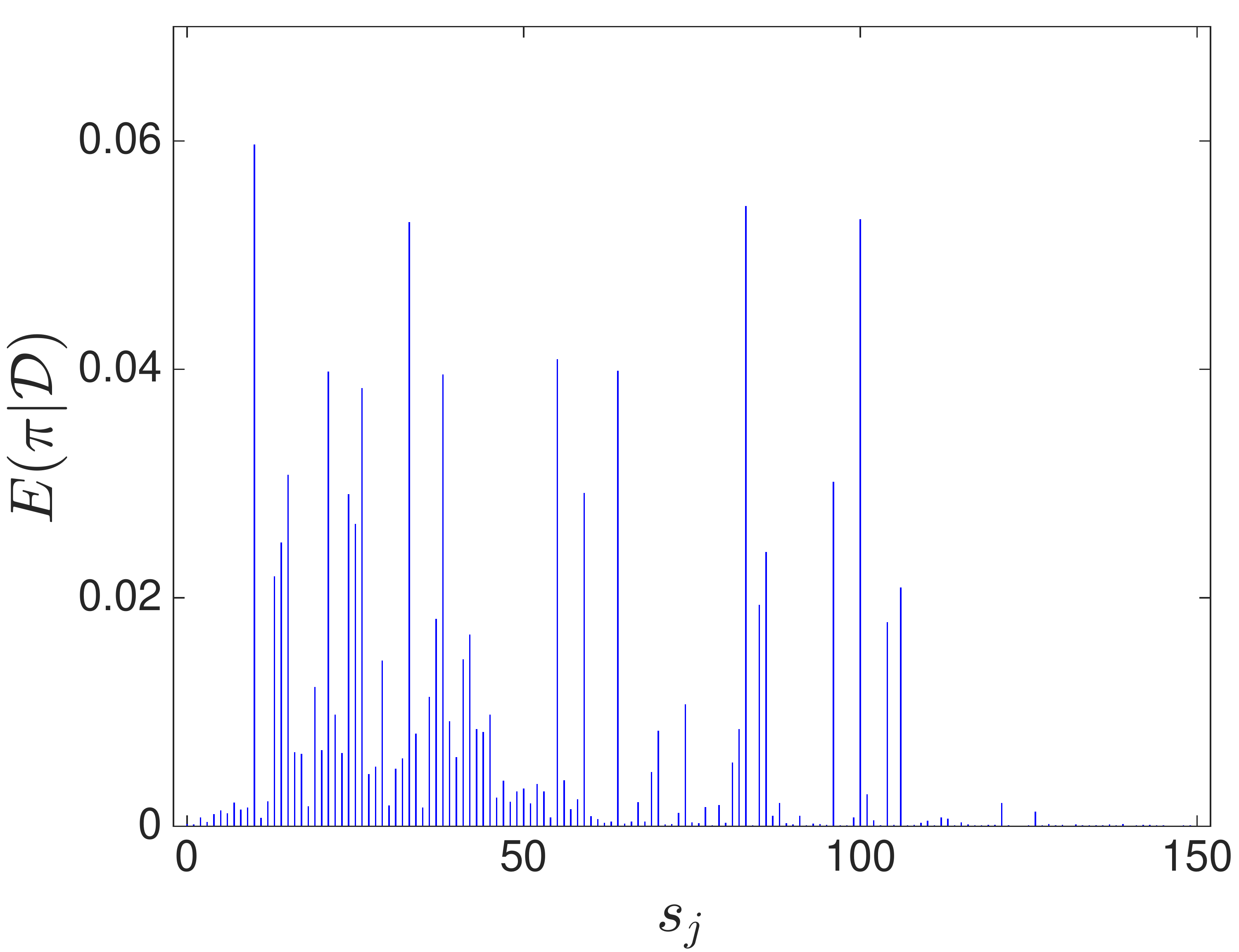}
\end{center}
\end{minipage}
\caption{{\protect\footnotesize The left hand side shows the posterior
distribution of the population probabilities of the quantiles }$\protect\pi %
_{j}=P(\protect\beta =s_{j})${\protect\footnotesize , }$\protect\tau =0.30$%
{\protect\footnotesize . The right hand side shows the corresponding result
for }$\protect\tau =0.90$. }
\label{fig:two values of tau}
\end{figure}

Figure \ref{fig:two values of tau} shows the common mixing distribution $%
\mathrm{E}\left( \pi |\mathcal{D}\right) $ for two quantile levels $\tau
=0.30$ and $\tau =0.90$. Notice, of course, how different they are, with a
great deal of mass on low scores when $\tau =0.30$ and vastly more scatter
for $\tau =0.90$. This is because even the very best batsmen fail with a
substantial probability, frequently recording very low scores. In the right
hand tail, the difference between the skill levels of the players is much
more stark, with enormous scatter.

\begin{figure}[th]
\centering
\includegraphics[width=14cm]{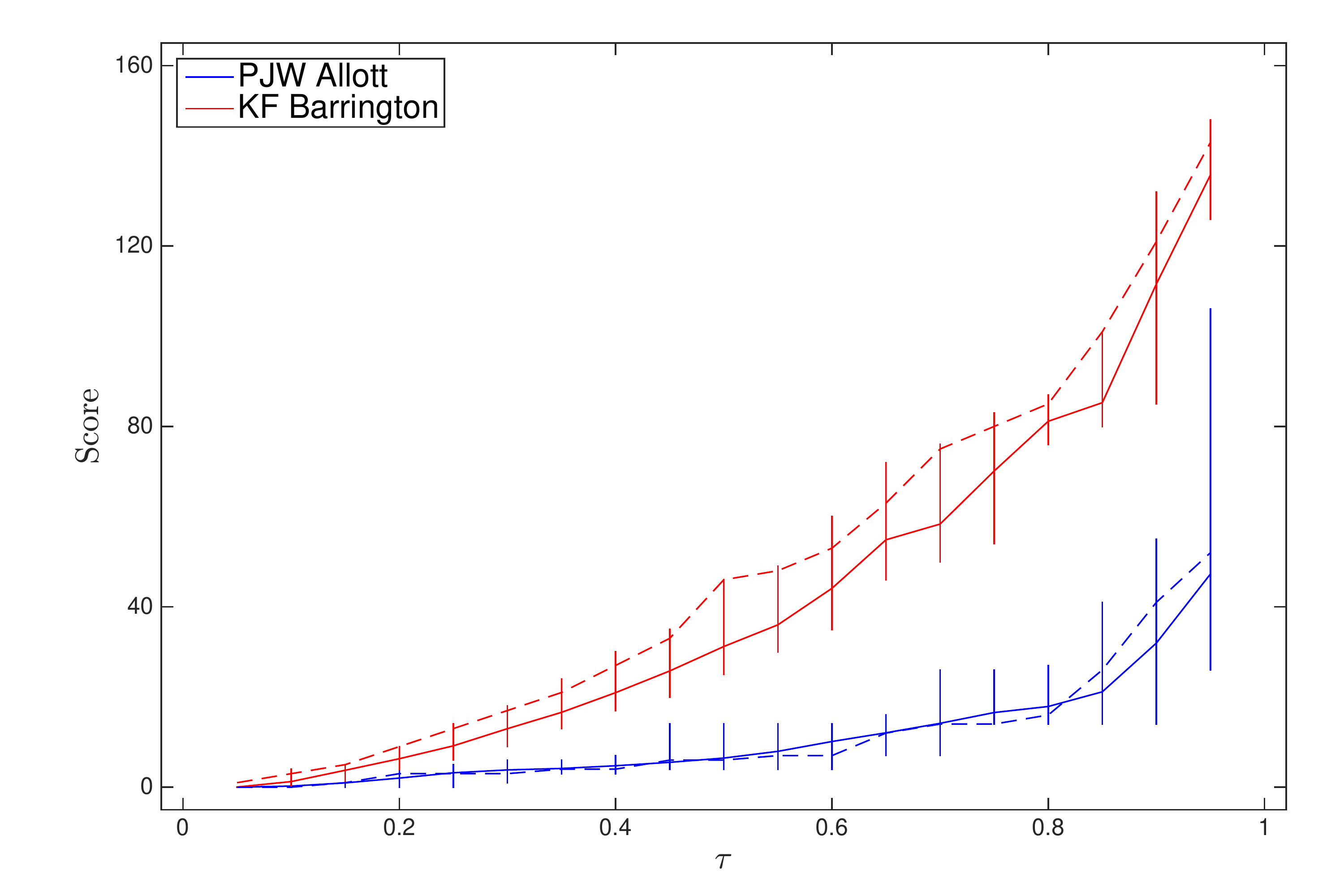}
\caption{{\protect\footnotesize The pointwise estimated quantile function
for two cricketers: P.J.W. Allott and K.F. Barrington. These calculations
ignore the impact of censoring. Horizonal lines denote 90\% posterior
intervals with 5\% in each tail. The curve for Allott uses his 18 innings,
Barrington had 131 innings. \ }}
\label{fig:quantF}
\end{figure}

We now turn to individual players. The dashed blue line in the left hand
side of Figure \ref{fig:quantF} shows the empirical quantile function for
P.J.W. Allott, while also plotted using a blue full line is the associated
Bayesian quantile function $\mathrm{E}(\beta _{\tau }|\mathcal{D})$. The
results are computed for $\tau \in \left\{ 0.01,0.2,...,0.99\right\} $. The
Bayesian function also shows a central $90$\% interval around the estimate.

The right hand side shows the same object but for K.F. Barrington, who
tended to score very highly and also played a great deal (his $n_{i}$ is
around 8 times larger than Allott's). We can see in both players' cases the
lower quantiles are very precisely estimated and not very different, but at
higher quantile levels the uncertainty is material and the differences in
level stretch out. Further, at these higher levels the 90\% intervals are
typically skewed, with a longer right hand tail. \ 

The Bayesian quantile functions seem shrunk more for Barrington, which looks
odd as Allott has a smaller sample size. But Barrington has typically much
higher scores (and so more variable) and so his quantiles are intrinsically
harder to estimate and so are more strongly shrunk. His exceptionalism is
reduced by the shrinkage.\newline

For a moment we now leave the cricket example. We should note that we have
ignored the fact some innings were not completed and marked
\textquotedblleft not out\textquotedblright, a form of censoring. We now
develop methods to overcome this deficiency. \ \ 

\section{Truncated data\label{sect:truncate}}

\subsection{Censored data}

Here we show how this methodology can be extended to models with truncated
data. The probabilistic aspect of the model is unaltered. We assume the
support is sorted and known to be $\mathcal{S}$, and $\Pr (Z=s_{j}|\theta
)=\theta _{j}$. \ However, in addition to some fully observed data, $%
\mathcal{D}_{1}=\{z_{1},...,z_{N}\}$, there exist $N^{\prime }$ additional
data, $\mathcal{D}_{2}=\{s_{l_{i}},...,s_{l_{N^{\prime }}}\}$, which we know
has been right truncated. We assume the non-truncated versions of the data
are independent over $i$, such that $U_{i}\geq s_{l_{i}}$, $1\leq i\leq
N^{\prime }$, $U_{i}\in \mathcal{S}$, $\Pr (U=s_{j}|\theta )=\theta _{j}$.
We write $\mathcal{U=}\{U_{1},...,U_{N^{\prime }}\}$. Therefore our data is $%
\mathcal{D}=\mathcal{D}_{1}\bigcup \mathcal{D}_{2}$.

Inference on $(\beta ,\pi )$ is carried out by augmenting it with $U$, and
employing a Gibbs sampler in order to draw from $p(\beta ,\pi ,U|\mathcal{D}%
) $.

\subsection{Computational aspects}

We implement this by Gibbs sampling, adding a first step to Algorithm 1. \ 

\begin{itemize}
\item \textbf{Algorithm 2: }$\beta ,\pi ,U|\mathcal{D}$\textbf{\ Gibbs
sampler\ }\ 
\end{itemize}

\begin{enumerate}
\item Sample $\Pr (U|\beta ,\mathcal{D},\pi ).$

\item Sample $\Pr (\beta |\mathcal{D},U,\pi ).$

\item Sample $f(\pi |\beta )$, returning to $1$. \ \ 
\end{enumerate}

Sampling from $U|\left( \beta =s_{k},\mathcal{D}\right) $ is not standard,
but is also not difficult. We carry this out through data augmentation:

\begin{enumerate}
\item Sampling from $\Pr (U|\beta ,\mathcal{D},\pi )$ by,

\begin{enumerate}
\item Sample $\theta |\left(\beta =s_{k},\mathcal{D}\right) \mathcal{\sim }%
D_{J}(\alpha +\mathbf{n},k)$.

\item Sample $U|\left( \beta =s_{k},\theta ,\mathcal{D}\right)$.
\end{enumerate}
\end{enumerate}

Step 1(b) is straightforward, while 1(a) is a truncated Dirichlet defined in
(\ref{truncated dirichlet}). The Appendix \ref{sect:constrained dirichlet}
shows how to simulate from $D_{J}(\alpha ,k)$ exactly. As a side remark, it
is tempting to sample $\theta |U,\mathcal{D}$ and $U|\theta ,\mathcal{D}$
but this fails in practice; the reasons for this are described in detail in
Appendix \ref{sect:failed simulator}.

\subsection{A Bayesian bootstrap for the censored data}

A Bayesian bootstrap algorithm can be developed to deal with the censored
data (however its extension to hierarchical model is not straightforward,
since priors on $\beta$ can not be incorporated in this algorithm).
Independent draws from the Bayesian bootstrap posterior distribution can be
obtained by the following algorithm.

\begin{itemize}
\item \textbf{Algorithm 3: Bayesian bootstrap with censored data}
\end{itemize}

\begin{enumerate}
\item Draw $\theta ^{\ast }\sim \text{Dirichlet}(\alpha +\mathbf{n})$.

\item For $1\leq i\leq N^{\prime }$, draw $U_{i}$ from $%
\{s_{l_{i}},...,s_{J}\}$, with probability $\Pr (U_{i}=s_{j})=\frac{\theta
_{j}^{\ast }}{\sum_{k=l_{i}}^{J}\theta _{k}^{\ast }}$, and set $%
n_{j}^{\prime }=\sum_{1}^{N^{\prime }}1(U_{i}=s_{j})$, and $\mathbf{n}%
^{\prime }=(n_{1}^{\prime },...,n_{J}^{\prime })$.

\item Draw $\theta \sim \text{Dirichlet}(\alpha +\mathbf{n}+\mathbf{n}%
^{\prime })$. Set $\beta =t(\theta )$. Go to $1$.
\end{enumerate}


\subsection{Returning to cricket: the impact of not outs}

In cricket scores at least 9\% of scores in each innings must be not out, so
right censoring is important statistically. Not outs are particularly
important for weaker batsmen who are often left not out at the end of the
team's innings. In Section \ref{sect:cricket} we ignored this feature of
batting and here we return to it to correct the results. \ 
\begin{figure}[th]
\begin{center}
\includegraphics[width=13cm]{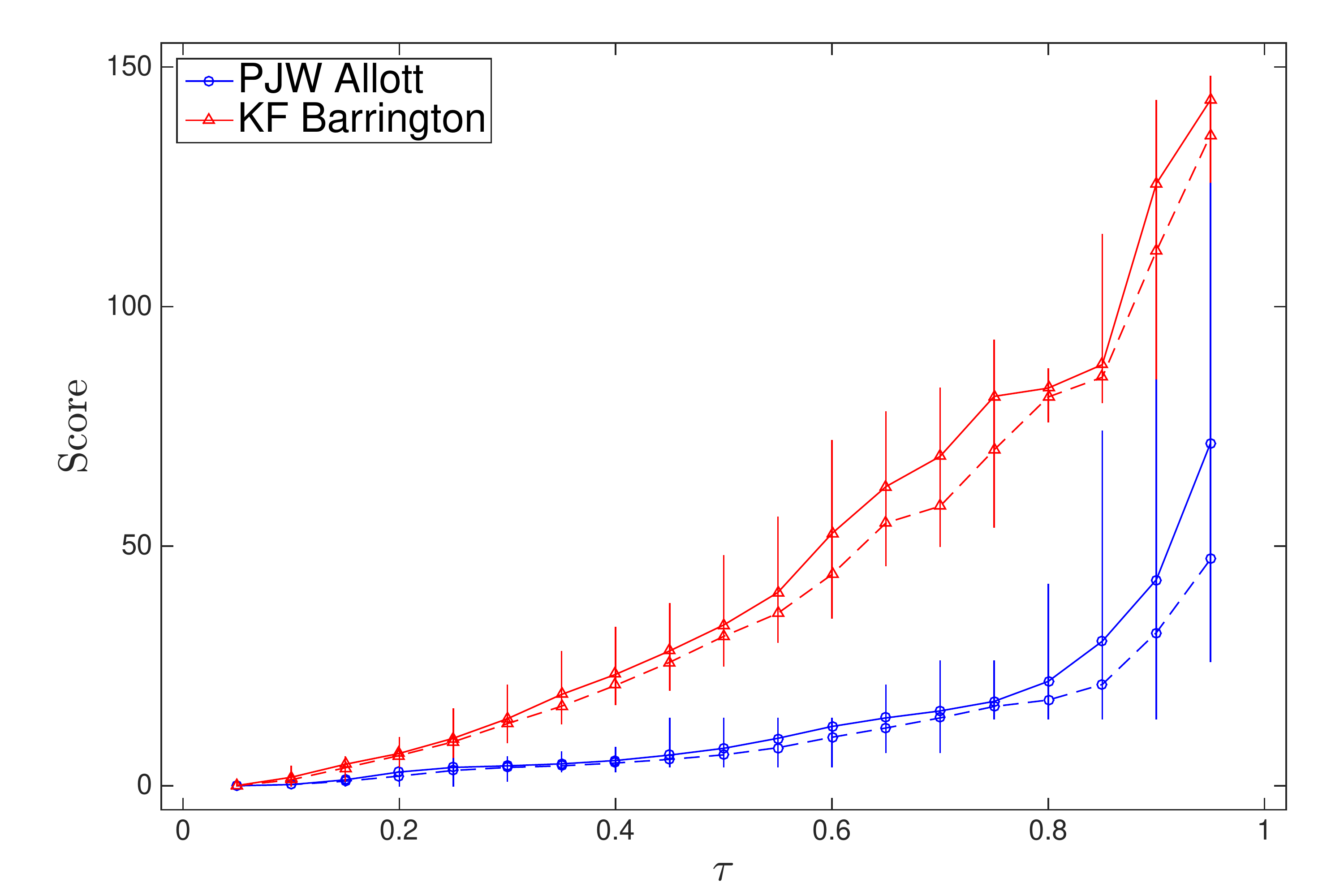}
\end{center}
\caption{{\protect\footnotesize The censored-adjusted pointwise estimated
quantile function for two cricketers: P.J.W. Allott and K.F. Barrington. The
solid lines are the the estimates with the censored observations, and the
dashed lines are obtained by ignoring that they are censored data. Horizonal
lines denote 90\% posterior intervals with 5\% in each tail. The curve for
Allott uses his 18 innings, Barrington had 131 innings.}}
\label{Fig:QE_Example11}
\end{figure}

Figure \ref{Fig:QE_Example11} shows the estimated pointwise quantile
function for Barrington and Allott, taking into account the not outs. \ Both
are shifted upwards, particularly Allott in the right hand tail. However,
Allott's right hand tail is not precisely estimated. \ 
\begin{table}[!ht]
\centering{\footnotesize \ 
\begin{tabular}{l||rrr||rrr|rrr}
& \multicolumn{3}{|c||}{} & \multicolumn{6}{|c}{Ignoring censoring in
analysis} \\ \hline
& \multicolumn{3}{|c||}{Bayesian} & \multicolumn{3}{|c}{Bayesian} & 
\multicolumn{3}{|c}{Empirical} \\ 
Batsman & $\widetilde{\beta }_{1/2}$ & $Q_{5}$ & $Q_{95}$ & $\widetilde{%
\beta }_{1/2}$ & $Q_{5}$ & $Q_{95}$ & $n_i$ & $n_i^{\prime }$ & $\widehat{%
\beta }_{1/2}$ \\ \hline
A Khan & 14.7 & 4 & 30 & 12.7 & 2 & 27 & 0 & 0 & -- \\ 
ACS Pigott & 9.7 & 4 & 27 & 7.8 & 4 & 19 & 2 & 1 & 4 \\ 
A McGrath & 14.3 & 4 & 31 & 13 & 4 & 27 & 5 & 0 & 34 \\ 
AJ Hollioake & 5.5 & 4 & 14 & 4.7 & 2 & 12 & 6 & 0 & 2 \\ 
JB Mortimore & 16.4 & 9 & 20 & 16.7 & 9 & 19 & 12 & 2 & 11 \\ 
DS Steele & 19.4 & 7 & 37 & 18.8 & 7 & 35 & 16 & 0 & 42 \\ 
PJW Allott & 7.8 & 4 & 14 & 6.7 & 4 & 14 & 18 & 3 & 6 \\ 
JC Buttler & 17.7 & 13 & 27 & 17.1 & 13 & 27 & 20 & 3 & 13 \\ 
W Larkins & 12.6 & 7 & 27 & 13.3 & 7 & 25 & 25 & 1 & 11 \\ 
NG Cowans & 5.9 & 4 & 10 & 4.1 & 3 & 7 & 29 & 7 & 3 \\ 
JK Lever & 6.7 & 4 & 11 & 5.3 & 4 & 8.5 & 31 & 5 & 6 \\ 
M Hendrick & 6.3 & 4 & 10 & 3.4 & 1 & 5 & 35 & 15 & 2 \\ 
DR Pringle & 8.3 & 7 & 10 & 7.7 & 4 & 9 & 50 & 4 & 8 \\ 
C White & 13.1 & 8 & 19 & 11.7 & 7 & 19 & 50 & 7 & 10 \\ 
GO Jones & 17.1 & 10 & 22 & 15.9 & 10 & 19 & 53 & 4 & 14 \\ 
CJ Tavare & 17.4 & 12.5 & 25 & 17.3 & 13 & 25 & 56 & 2 & 22 \\ 
PCR Tufnell & 4.8 & 1 & 9 & 1.2 & 0 & 2 & 59 & 29 & 1 \\ 
MS Panesar & 4 & 4 & 4 & 1.6 & 0 & 4 & 64 & 21 & 1 \\ 
CM Old & 8.9 & 7 & 13 & 8.4 & 7 & 11 & 66 & 9 & 9 \\ 
JA Snow & 7.9 & 4 & 9 & 5.5 & 4 & 8 & 71 & 14 & 6 \\ 
DW Randall & 14.6 & 9 & 19 & 14.5 & 10 & 19 & 79 & 5 & 15 \\ 
RC Russell & 16.2 & 9.5 & 24 & 14.6 & 12 & 20 & 86 & 16 & 15 \\ 
MR Ramprakash & 18.8 & 14 & 21 & 18.6 & 14 & 21 & 92 & 6 & 19 \\ 
PD Collingwood & 25 & 19 & 30 & 23.3 & 19 & 28 & 115 & 10 & 25 \\ 
RGD Willis & 8.5 & 7 & 10 & 4.1 & 4 & 5 & 128 & 55 & 5 \\ 
KF Barrington & 33.7 & 27 & 48 & 31 & 25 & 46 & 131 & 15 & 46 \\ 
APE Knott & 18.9 & 14 & 27 & 17.5 & 13 & 24 & 149 & 15 & 19 \\ 
IT Botham & 20.6 & 15 & 27 & 20.7 & 15 & 27 & 161 & 6 & 21 \\ 
DI Gower & 27.5 & 26 & 32 & 27 & 25 & 28 & 204 & 18 & 27 \\ 
AJ Stewart & 26.3 & 19 & 29.5 & 25.6 & 19 & 28 & 235 & 21 & 27%
\end{tabular}%
}
\caption{{\protect\footnotesize Estimated median batting scores. Sample
median is compared with two Bayesian estimators, where }$\protect\widetilde{%
\protect\beta }_{1/2}${\protect\footnotesize \ = }$E(\protect\beta _{1/2}|D)$%
{\protect\footnotesize . }$n_{i}${\protect\footnotesize \ is the number of
innings, }$n_{i}^{\prime }${\protect\footnotesize \ denotes the number of
not outs which are treated as right censored data and }$\protect\widehat{%
\protect\beta }_{1/2}${\protect\footnotesize \ is the empirical median. In
the first model the not outs are assumed to be right censored observations.
In the second model they are treated as if they were completed innings. }$%
Q_{5}${\protect\footnotesize \ denotes the estimated 5\% point on the
relevant posterior distribution. }}
\label{tab:CensoringCorrect}
\end{table}

Table \ref{tab:CensoringCorrect} shows the Bayesian results for our selected
30 players, updating Table \ref{tab:cricketSubpop} to reflect the role of
right censoring. Here $n_{i}^{\prime }$ denotes the number of not out, that
is right censored innings, the player had. In many cases this is between
10\% and 20\% of the innings, but for some players it is far higher. R.G.S.
Willis is the leading example, who had 55 not outs of 128 innings. A leading
bowler, he usually batted towards the end of innings and was often left not
out. His posterior mean of the median is inflated greatly by the statistical
treatment of censoring. Further, the interval between $Q_{5}$ and $Q_{95}$
is widened substantially. Other players are hardly affected, e.g. M.R.
Ramprakash, who had 6 not outs in 92 innings.

Table \ref{tab:CensoringCorrect} shows a ranking of players by the mean of
the posteriors of the quantiles, at three different levels of quantiles.
This shows how the rankings change greatly with the quantile level. For
small levels, we can think of this as being about consistency. For the
median it is about typical performance. For the 90\% quantile this is about
upside potential to bat long. A remarkable result is J.B. Bolus who has a
very high $\beta _{0.30}$ quantile. His career innings were the following:
14, 43, 33, 15, 88, 22, 25, 57, 39, 35, 58, 67. He only played for a single
year, but never really failed in a single inning. However he never managed
to put together a very long memorable innings and this meant his Test career
was cut short by the team selectors. They seem to not so highly value
reliability. \ \ 
\begin{table}[tbp]
\centering{\footnotesize \ 
\begin{tabular}{r||lrrr||lrrr|lrrr}
& \multicolumn{4}{c}{0.3 quantile} & \multicolumn{4}{c}{0.5 quantile} & 
\multicolumn{4}{c}{0.9 quantile} \\ \hline
rank & Batsman & $\widetilde{\beta }_{0.3}$ & Q$_{5}$ & Q$_{95}$ & Batsman & 
$\widetilde{\beta }_{0.5}$ & Q$_{5}$ & Q$_{95}$ & Batsman & $\widetilde{%
\beta }_{0.9}$ & Q$_{5}$ & Q$_{95}$ \\ \hline
1 & JB Bolus & 16.6 & 4 & 33 & KF Barrington & 33.7 & 27 & 48 & KF Barrington
& 121.3 & 101 & 143 \\ 
2 & KF Barrington & 14.0 & 9 & 21 & KP Pietersen & 30.5 & 26 & 34 & IR Bell
& 116.8 & 109 & 121 \\ 
3 & DI Gower & 13.1 & 11 & 16 & JH Edrich & 29.5 & 22 & 35 & GP Thorpe & 
115.8 & 94 & 119 \\ 
4 & AN Cook & 12.9 & 11 & 13 & G Boycott & 29.1 & 23 & 35 & PH Parfitt & 
115.6 & 86 & 121 \\ 
5 & ER Dexter & 12.8 & 10 & 16 & ER Dexter & 28.5 & 27 & 32 & IJL Trott & 
111.2 & 64 & 121 \\ 
6 & G Boycott & 12.6 & 10 & 13 & ME Trescothick & 28.4 & 24 & 32 & MC Cowdrey
& 111.1 & 96 & 119 \\ 
7 & GA Gooch & 12.6 & 10 & 13 & BL D'Oliveira & 28.2 & 23 & 32 & G Boycott & 
109.7 & 106 & 116 \\ 
8 & KP Pietersen & 12.6 & 9 & 14 & AJ Strauss & 28.1 & 25 & 32 & DL Amiss & 
106.9 & 64 & 119 \\ 
9 & RW Barber & 12.5 & 6 & 13 & R Subba Row & 27.9 & 22 & 32 & AN Cook & 
106.7 & 96 & 118 \\ 
10 & AJ Strauss & 12.5 & 9 & 14 & DI Gower & 27.5 & 26 & 32 & MP Vaughan & 
106.3 & 100 & 115 \\ 
11 & G Pullar & 12.4 & 9 & 14 & MC Cowdrey & 27.3 & 23 & 32 & ME Trescothick
& 105.4 & 90 & 113 \\ 
12 & ME Trescothick & 12.4 & 9 & 14 & AW Greig & 27.2 & 19 & 32 & KP
Pietersen & 105.2 & 96 & 119 \\ 
13 & MP Vaughan & 12.4 & 9 & 13 & AN Cook & 27.2 & 22 & 32 & AJ Strauss & 
105.0 & 83 & 112 \\ 
14 & MC Cowdrey & 12.3 & 9 & 13 & GA Gooch & 27.1 & 22 & 30 & AW Greig & 
102.8 & 96 & 110 \\ 
15 & JE Root & 12.2 & 6 & 13 & JB Bolus & 27.1 & 15 & 36 & N Hussain & 102.5
& 85 & 109 \\ 
16 & R Subba Row & 12.1 & 8 & 13 & GP Thorpe & 27.1 & 19 & 32 & CT Radley & 
102.4 & 59 & 106 \\ 
17 & RA Smith & 12.1 & 8 & 13 & IJL Trott & 26.7 & 19 & 35 & JE Root & 102.1
& 83 & 130 \\ 
18 & JM Parks & 12.0 & 7 & 14 & AJ Stewart & 26.3 & 19 & 29 & DI Gower & 
101.8 & 85 & 106 \\ 
19 & JG Binks & 12.0 & 6 & 13 & PH Parfitt & 25.8 & 18 & 32 & AJ Lamb & 101.4
& 83 & 119 \\ 
20 & GP Thorpe & 11.8 & 9 & 13 & MP Vaughan & 25.7 & 19 & 32 & DS Steele & 
100.8 & 64 & 106%
\end{tabular}%
}
\caption{Best $20$ players ranked based on the mean of the posteriors of the
quantiles, at three different levels of quantiles. }
\label{tab:RankingofPlayers1}
\end{table}

Again K.F. Barrington is the standout batsman. He is very strong at all the
different quantiles. Notice though he still had a 30\% chance of scoring 14
or less --- which would be regarded by many cricket watchers as a failure.
But once his innings was established his record was remarkably strong,
typically playing long innings. \

\section{Conclusions \label{section:conclusions}}

In this paper we provide a Bayesian analysis of quantiles by embedding the
quantile problem in a larger inference challenge. This delivers quite simple
ways of performing inference on a single quantile. The frequentist
performance of our methods are similar to that of the bootstrap. \ 

We extend the framework to introduce a hierarchical quantile model, where
each subpopulation's distribution is modeled nonparametrically but linked
through a nonparametric mixing distribution placed on the quantile. This
allows non-linear shrinkage, adjusting to skewed and sparse data in an
automatic manner. \ 

This approach is illustrated by the analysis of a large database from sports
statistics of 300 Test cricketers. Each person's batting performance is
modeled nonparametrically and separately, but linked through a quantile
which is drawn from a common distribution. This allows us to shrink each
cricketer's performance -- a particular advantage in cases where the careers
are very short.

The modeling approach is extended to allow for truncated data. This is
implemented by using simulation based inference. Again this set is
illustrated in practice by looking at not outs in batting innings, where we
think of the data as right censored.

\baselineskip=14pt

\bibliographystyle{chicago}
\bibliography{REZA}

\baselineskip=20pt

\appendix%

\section{Appendix}

\subsection{Proof or Proposition \protect\ref{Prop:uniqueness of beta}}

As $\Psi (b,\theta )$ is a convex function it has a unique minimizer on $%
\mathbb{R}$, or its optimal set is a closed interval, $[\beta _{l},\beta
_{u}]$, where $s_{1}\leq \beta _{l}<\beta _{u}\leq s_{J}$ (\cite%
{Rockafellar(70)}). In the latter case, there exist $\beta _{m}\in \lbrack
\beta _{l},\beta _{u}]\backslash \mathcal{S}$, at which $\frac{\partial \Psi
(b,\theta )}{\partial b}$ exists and is equal to zero, 
\begin{equation*}
\left. \frac{\partial \Psi (b,\theta )}{\partial b}\right\vert _{b=\beta
_{m}}=(1-\tau )\tau ^{\prime }-\tau (1-\tau ^{\prime })=0
\end{equation*}%
where $\tau ^{\prime }=\sum_{j=1}^{k}\theta _{j}$, and $k=\max
\{j;s_{j}<\beta _{m}\}$. For a specific value of $\beta _{m}$, this equality
holds if $\tau ^{\prime }=\tau $, that means, $\sum_{j=1}^{k}\theta
_{j}=\tau $. This implies that the minimizer of $\Psi (b,\theta )$ is not
unique if and only if $\tau \in \{\theta _{1},\theta _{1}+\theta
_{2},....,\theta _{1}+\cdots +\theta _{J-1}\}$, and this is a zero measure
event if $\theta $ is non-singular.

Now assume $\beta ^{\ast }$ is the unique minimizer of $\Psi (b,\theta )$
for $\theta =\theta ^{\ast }$. This implies that the directional derivatives
of $\Psi (b,\theta )$ are strictly positive at the optimal point, $\nabla
_{v}\Psi (\beta ^{\ast },\theta ^{\ast })>0$, for$\quad v\in \{-1,1\}$.
However, $\nabla _{v}\Psi (\theta ,b)$ is an affine (and therefore a
continuous) function of $\theta $, 
\begin{equation*}
\nabla _{v}\Psi (b,\theta )=\sum_{j=1}^{J}(1-\tau )\theta
_{j}v1(s_{j}<b)-\tau \theta _{j}v1(s_{j}>b)+\theta _{j}\rho _{\tau
}(-v)1(s_{j}=b),
\end{equation*}%
therefore there exists an open ball centered at $\theta ^{\ast }$ with
radius $\delta >0$, $\mathcal{B}_{\delta }(\theta ^{\ast })$, in such a way
that, $\forall v\in \{-1,1\}$, $\forall \theta \in \mathcal{B}_{\delta
}(\theta ^{\ast })$, $\nabla _{v}\Psi (\beta ^{\ast },\theta )>0$. Hence,
for any $\theta \in \mathcal{B}_{\delta }(\theta ^{\ast })$, the objective
function $\Psi (b,\theta )$ has a unique minimizer at $\beta ^{\ast }$: $%
\forall \theta \in \mathcal{B}_{\delta }(\theta ^{\ast })$, $\beta ^{\ast }=%
\underset{b}{\func{argmin}}\ \Psi (b,\theta )$, and this implies, $\left.
\partial \beta /\partial \theta ^{\prime }\right\vert _{\theta =\theta
^{\ast }}=0$.

\subsection{Proof of Proposition \protect\ref{Prop Dirich c}}

For $k=1$, $\mathcal{A}_{1}=\{\theta ;\theta _{1}>\tau \}$, therefore, $%
c_{1}=\Pr (\theta \in \mathcal{A}_{1})=\Pr (\theta _{1}>\tau )$. For $2\leq
k\leq J-1$, we have, 
\begin{equation*}
\mathcal{A}_{k}=\left\{ \theta ;\sum_{j=1}^{k-1}\theta _{j}<\tau ,\ \ \text{%
and}\ \ \sum_{j=1}^{k}\theta _{j}>\tau \right\} =\left\{ \theta
;\sum_{j=1}^{k-1}\theta _{j}<\tau \right\} \backslash \left\{ \theta
;\sum_{j=1}^{k}\theta _{j}\leq \tau \right\} .
\end{equation*}%
Since $\left\{ \theta ;\sum_{j=1}^{k}\theta _{j}\leq \tau \right\} \subset
\left\{ \theta ;\sum_{j=1}^{k-1}\theta _{j}<\tau \right\} $, then, $%
c_{k}=\Pr (\theta \in \mathcal{A}_{k})=\Pr \left( \sum_{j=1}^{k-1}\theta
_{j}<\tau \right) -\Pr \left( \sum_{j=1}^{k}\theta _{j}<\tau \right) $.
Finally, for $k=J$, $\mathcal{A}_{J}=\left\{ \theta ;\sum_{j=1}\theta
_{j}<\tau \right\} $, and so $c_{J}=\Pr \left( \sum_{j=1}\theta _{j}<\tau
\right) $.

\subsection{Proof of Proposition \protect\ref{Prop of posterior for beta}}

In the Dirichlet case the posterior distribution of $\beta ,\theta $ will
be, 
\begin{equation*}
p(\beta =s_{k(\theta )},\theta |\mathcal{D})\propto \left(
\prod_{j=1}^{J}\theta _{j}^{n_{j}}\right) \frac{b_{k(\theta )}}{c_{k(\theta
)}(\alpha )}f_{D}(\theta ;\alpha )=\frac{1}{c(\alpha +n)}\frac{b_{k(\theta )}%
}{c_{k(\theta )}(\alpha )}f_{D}(\theta ;\alpha +n),
\end{equation*}%
where $n=(n_{1},...,n_{J})$ and $c(\alpha +n)=\int_{\Delta}f_{D}(\theta
;\alpha +n)\mathrm{d}\theta $. The right hand side integrates to $C(\alpha
,n)=\sum_{k=1}^{J}\frac{c_{k}(\alpha +n)}{c_{k}(\alpha )}b_{k}$, hence the
normalized posterior is $p(\beta =s_{k(\theta )},\theta |\mathcal{D}%
)=m_{k(\theta )}f_{D}(\theta ;\alpha +n)$, where $m_{k}=\frac{1}{C(\alpha ,n)%
}\frac{b_{k}}{c_{k}(\alpha )}$.

The posterior distribution of $\beta $ can be found analytically, recalling
that for quantiles the area formula implies $p(\theta |\mathcal{D})=p(\beta
,\theta |\mathcal{D})$, as $\Pr (\beta =s_{k}|\mathcal{D})=\int_{\mathcal{A}%
_{k}}p(\theta |\mathcal{D})\mathrm{d}\theta =\frac{1}{C(\alpha ,n)}\frac{%
c_{k}(\alpha +n)}{c_{k}(\alpha )}b_{k}$. 

\subsection{Proof of Proposition \protect\ref{prop small alpha}}

Note that, 
\begin{eqnarray*}
\underset{\alpha ,\beta \rightarrow 0}{\lim }B(\alpha ,\beta ) &=&\underset{%
\alpha ,\beta \rightarrow 0}{\lim }\frac{\Gamma (\alpha )\Gamma (\beta )}{%
\Gamma (\alpha +\beta )}=\underset{\alpha ,\beta \rightarrow 0}{\lim }\frac{%
\Gamma (\alpha +1)\Gamma (\beta +1)}{\Gamma (\alpha +\beta +1)}\frac{\alpha
+\beta }{\alpha \beta }=\underset{\alpha ,\beta \rightarrow 0}{\lim }\frac{%
\alpha +\beta }{\alpha \beta }, \\
\underset{\alpha ,\beta \rightarrow 0}{\lim }\alpha B(\tau ;\alpha ,\beta )
&=&\underset{\alpha ,\beta \rightarrow 0}{\lim }\alpha \frac{\tau ^{\alpha
}(1-\tau )^{\beta }}{\alpha }\left( 1+\frac{\alpha +\beta }{\alpha +1}\tau +%
\frac{(\alpha +\beta )(\alpha +\beta +1)}{(\alpha +1)(\alpha +2)}\tau
^{2}+\cdots \right) =1.
\end{eqnarray*}%
%
%
Hence, 
$\underset{\alpha ,\beta \rightarrow 0}{\lim }I_{\tau }(\alpha ,\beta
)=\beta /\left( \alpha +\beta \right) $. Assume all the elements of $\alpha $
goes to $0$ at the same rate, 
\begin{eqnarray*}
c_{k}(\alpha ) &\rightarrow &\underset{\alpha _{k}\downarrow 0}{\lim }\
I_{\tau }(\alpha _{k-1}^{+},\alpha _{J}^{+}-\alpha _{k-1}^{+})-I_{\tau
}(\alpha _{k}^{+},\alpha _{J}^{+}-\alpha _{k}^{+}) \\
&\rightarrow &\underset{\varepsilon \downarrow 0}{\lim }\ I_{\tau
}((k-1)\varepsilon ,(J-k+1)\varepsilon )-I_{\tau }(k\varepsilon
,(J-k)\varepsilon )=\frac{J-k+1}{J}-\frac{J-k}{J}=\frac{1}{J}.
\end{eqnarray*}%
As $\alpha \downarrow 0$, so $\theta _{k}^{+}\overset{L}{\rightarrow }\text{%
Beta}(n_{k}^{+},n-n_{k}^{+})$. Now using that limit, $\Pr (\theta
_{k}^{+}<\tau )=I_{\tau }(n_{k}^{+},n-n_{k}^{+})$. For $\tau \in (0,1)$, and
positive integers $k$ and $n$, 
$I_{\tau }(k,n)=1-F_{\text{B}}(k-1;n+k-1,\tau )$. 
So, for $2\leq k\leq J-1$, 
\begin{eqnarray*}
c_{k} &=&\Pr (\theta _{k-1}^{+}<\tau )-\Pr (\theta _{k}^{+}<\tau )=I_{\tau
}(n_{k-1}^{+},n-n_{k-1}^{+})-I_{\tau }(n_{k}^{+},n-n_{k}^{+}) \\
&=&F_{\text{B}}(n_{k}^{+}-1;n-1,\tau )-F_{\text{B}}(n_{k-1}^{+}-1;n-1,\tau
)=\sum_{k=n_{k-1}^{+}}^{n_{k}^{+}-1}f_{\text{B}}(k;n-1,\tau ), \\
c_{1} &=&1-\Pr (\theta _{1}^{+}<\tau )=F_{\text{B}}(n_{1}^{+}-1;n-1,\tau
)=\sum_{k=0}^{n_{1}^{+}-1}f_{\text{B}}(k;n-1,\tau
)=\sum_{k=n_{0}^{+}}^{n_{1}^{+}-1}f_{\text{B}}(k;n-1,\tau ) \\
c_{J} &=&\Pr (\theta _{J-1}^{+}<\tau )=1-F_{\text{B}}(n_{J-1}^{+}-1;n-1,\tau
)=\sum_{k=n_{J-1}^{+}}^{n-1}f_{\text{B}}(k;n-1,\tau
)=\sum_{k=n_{J-1}^{+}}^{n_{J}^{+}-1}f_{\text{B}}(k;n-1,\tau ),
\end{eqnarray*}%
where $n_{0}^{+}=0$. %
%
If there are no ties, n=J, and  $z_{1}=s_{1}<\cdots <z_{J}=s_{J}$, then the
result holds. \ 

\section{A basic simulator which does not work\label{sect:failed simulator}}

The $U=(U_{1},...,U_{N^{^{\prime }}})$, will be treated as missing data, and
the inference can be performed by sampling from $\pi ,\theta ,U|\mathcal{D}$%
. To do this we would need to sample from $p(\theta ,U|\mathcal{D})$.

One approach to sampling from this is using a Gibbs sampler.

\begin{itemize}
\item \textbf{Algorithm 2: }$\theta ,U|\mathcal{D}$\textbf{\ Gibbs sampler}
\end{itemize}

\begin{enumerate}
\item Draw from, $\theta |\mathcal{D},U\sim $Dirichlet$(\alpha +\mathbf{n}+%
\mathbf{n}^{\prime })$, $\mathbf{n}^{\prime }=(n_{1}^{\prime
},...,n_{J}^{\prime })$, $n_{j}^{^{\prime }}=\sum_{i=1}^{N^{\prime
}}1(U_{i}=s_{j})$. 

\item Draw from $p(U|\theta ,\mathcal{D})=p(U|\theta )$, $\Pr
(U_{i}=s_{j}|\theta )=\left\{ \theta _{j}/\sum_{k=l_{i}}^{J}\theta
_{k}\right\} 1_{\{s_{l_{i}},...,s_{J}\}}(s_{j})$, $i=1,2,...,N^{\prime }$.
\end{enumerate}

This Gibbs sampler sometimes performs well. However, if some of the missing
data is constrained to fall in a block with no other data, which we call
\textquotedblleft isolated missingness\textquotedblright , then there are
numerical difficulties. In the case of right censored data, a censored data
point is suffering from isolated missingness if $s_{l_{i}}>\underset{k}{%
\func{max}}\ \{z_{1},...,z_{N}\}$. Assume $\alpha$ is small and $U_{i}=s_{j}$%
. Then, for $\theta$ simulated in Step 1, with high probability we have $%
\theta _{j}/\sum_{k=l_{i}}^{J}\theta _{k}\simeq 1$, and so at Step 2 with
high probability $U_{i}=s_{j}$. The result is a highly correlated Gibbs
chain and this form of simulation is highly likely to fail.

\section{Constrained Dirichlet sampling\label{sect:constrained dirichlet}}

Assume $\theta \sim \text{Dirichlet}(\alpha )$, and consider simulating from 
$\theta |\beta =s_{k}$, for $k=1,...,J$. This is equivalent to simulating
from $\theta \sim $Dirichlet$(\alpha )1_{\mathcal{A}_{k}}(\theta ) $. Let $%
D_{J}(\alpha ,k)$ denote the $J-1$ dimensional Dirichlet distribution with
the parameters $\alpha =(\alpha _{1},...,\alpha _{J})$, and truncated to $%
\mathcal{A}_{k}$, for $k=1,...,J$, with the following density function: $%
p(\theta )=f_{D}(\theta ;\alpha )/c_{k}(\alpha )$. Below we show how we can
sample from $D_{3}(\alpha ,k)$. \ Once this is developed the generalization
to $J>3$ is straightforward.

\begin{itemize}
\item \textbf{Algorithm: sampling from }$D_{3}(\alpha ,k)$
\end{itemize}

\begin{enumerate}
\item $\mathcal{A}_{1}$: Draw from $\theta \sim \text{Dirichlet}(\alpha )1_{%
\mathcal{A}_{1}}(\theta )$, by: draw $\theta _{1}$ from $\text{Beta}(\alpha
_{1},\alpha _{2}+\alpha _{3})1_{(\tau ,1)}(\theta _{1})$; then draw $\theta
_{2}$ from $(1-\theta _{1})\text{Beta}(\alpha _{2},\alpha _{3})$.

\item $\mathcal{A}_{2}$: Draw from $\theta \sim \text{Dirichlet}(\alpha )1_{%
\mathcal{A}_{2}}(\theta )$ by: draw $\theta _{1}$ from $\text{Beta}(\alpha
_{1},\alpha _{2}+\alpha _{3})1_{(0,\tau )}(\theta _{1})$; then draw $\theta
_{2}$ from $(1-\theta _{1})\text{Beta}(\alpha _{2},\alpha _{3})$, until $%
\theta _{1}+\theta _{2}>\tau $. However, the rejection step could be
inefficient. Now 
\begin{equation*}
f(\theta _{1}|\theta \in \mathcal{A}_{2})=\frac{1}{c_{2}(\alpha )}%
f_{B}(\theta _{1};\alpha _{1},\alpha _{2}+\alpha _{3})\left[ 1-F_{B}\left( 
\frac{\tau -\theta _{1}}{1-\theta _{1}};\alpha _{2},\alpha _{3}\right) %
\right] 1_{(0,\tau )}(\theta _{1}),
\end{equation*}

Therefore we can instead use the more reliable alternative: draw $\theta
_{1} $ from $f(\theta _{1}|\theta \in \mathcal{A}_{2})$; draw $\theta _{2}$
from $(1-\theta _{1})\text{Beta}(\alpha _{2},\alpha _{3})1_{(\frac{\tau
-\theta _{1}}{1-\theta _{1}},1)}(\theta _{2})$.

\item $\mathcal{A}_{3}$: Draw from $\theta \sim \text{Dirichlet}(\alpha )1_{%
\mathcal{A}_{3}}(\theta )$ by drawing $\theta _{3}$ from $\text{Beta}(\alpha
_{3},\alpha _{1}+\alpha _{2})1_{(1-\tau ,1)}(\theta _{3})$; then draw $%
\theta _{1}$ from $(1-\theta _{3})\text{Beta}(\alpha _{1},\alpha _{2})$.
\end{enumerate}



\begin{itemize}
\item \textbf{Algorithm: sampling from }$D_{J}(\alpha ,k)$\textbf{\ for }$%
J>3 $\textbf{.}
\end{itemize}

\begin{enumerate}
\item For $k=1$: draw $(\theta _{1},S,\theta _{J})$ from $D_{3}((\alpha
_{1},\alpha^{+}_{J-1}-\alpha^{+}_{1},\alpha _{J}),1)$; then draw $\frac1S
(\theta _{2},...,\theta _{J-1})$ from $\text{Dirichlet}(\alpha_{2}, ...,
\alpha_{J-1})$.

\item For $2\leq k\leq J-1$: draw $(S_{1},\theta _{k},S_{2})$ from $%
D_{3}((\alpha^{+}_{k-1},\alpha _{k},\alpha^{+}_{J}-\alpha^{+}_{k}),2)$; draw 
$\frac{1}{S_{1}} (\theta_{1},...,\theta _{k-1})$ from $\text{Dirichlet}%
(\alpha_{1}, ..., \alpha_{k-1})$; draw $\frac{1}{S_{2}}(\theta
_{k+1},...,\theta _{J})$ from $S_{2}\ \text{Dirichlet}(\alpha_{k+1}, ...,
\alpha_{J})$.

\item For $k=J$: draw $(S,\theta _{J-1},\theta _{J})$ from $%
D_{3}((\alpha^{+}_{J-2},\alpha _{J-1},\alpha _{J}),3)$; then draw $\frac1S
(\theta _{1},...,\theta_{J-2})$ from $\text{Dirichlet}(\alpha_{1}, ...,
\alpha_{J-2})$.
\end{enumerate}

\section{Sampling from truncated beta distribution}

Here we simulate $\theta $ from $\text{Beta}(\alpha ,\beta )$, truncated to $%
(L,U)$ interval, where $0\leq L<U\leq 1$ (assuming either $L\neq 0$ or $%
U\neq 1$). Inverse transform sampling will be numerically infeasible if $%
F_{B}(L;\alpha ,\beta )$ and $F_{B}(U;\alpha ,\beta )$ are both very close
to $0$ or $1$. We can distinguish $4$ cases. All other cases can be
transformed to one of the four cases by $\text{Beta}(\alpha ,\beta )\overset{%
d}{=}1-\text{Beta}(\beta ,\alpha )$.

\begin{itemize}
\item \textbf{Algorithm: }$\alpha <1$\textbf{, }$\beta <1$\textbf{, and }$%
U<1 $. Draws from $\text{Beta}(\alpha ,\beta )$ truncated to $(L,U)$:
\end{itemize}

\begin{enumerate}
\item Draw $u\sim \text{Uniform}(0,1)$, and set $z^{\ast }=\frac{1}{\alpha }%
\ln (L^{\alpha }+(U^{\alpha }-L^{\alpha })u)$.

\item Draw $v\sim \text{Uniform}(0,1)$; if $v\leq \left( \frac{1-e^{z^{\ast
}}}{1-U}\right) ^{\beta -1}$, set $z=z^{\ast }$, otherwise go to step (1).

\item Return $\theta =e^{z}$.
\end{enumerate}

This is a rejection sampler for $z=\ln \theta ,$ with the proposal density $%
f_{Z}(z)=\frac{\alpha e^{\alpha z}}{U^{\alpha }-L^{\alpha }}\ 1_{(\log
L,\log U)}(z)$.

\begin{itemize}
\item \textbf{Algorithm: }$\alpha <1$\textbf{, }$\beta >1$\textbf{, and }$%
0<L $. Draws from $\text{Beta}(\alpha ,\beta )$ truncated to $(L,U)$:
\end{itemize}

\begin{enumerate}
\item Draw $u\sim \text{Uniform}(0,1)$, and set $\theta ^{\ast }=1-\left[
(1-L)^{\beta }-\left[ (1-L)^{\beta }-(1-U)^{\beta }\right] u\right] ^{\frac{1%
}{\beta }}$.

\item Draw $v\sim \text{Uniform}(0,1)$; if $v\leq \left( \frac{\theta ^{\ast
}}{L}\right) ^{\alpha -1}$, set $\theta =\theta ^{\ast }$, otherwise go to
step 1.
\end{enumerate}

This is a rejection sampler for $\theta ,$ with the proposal density $%
f_{Z}(z)=\frac{\beta }{(1-L)^{\beta }-(1-U)^{\beta }}(1-\theta )^{\beta -1}$.

\begin{itemize}
\item \textbf{Algorithm: }$\alpha >1$\textbf{, }$\beta <1$\textbf{.} If $%
\frac{\beta \ B(\alpha ,\beta )}{U^{\alpha -1}\left[ (1-L)^{\beta
}-(1-U)^{\beta }\right] }\leq 1$, then generate $\theta $ from $\text{B}%
(\alpha ,\beta )$, and accept if $L\leq \theta \leq U$. Otherwise, if $0<L$,
the following rejection algorithm is more efficient:
\end{itemize}

\begin{enumerate}
\item Draw $u\sim \text{Uniform}(0,1)$, and set $\theta ^{\ast }=1-\left[
(1-L)^{\beta }-\left[ (1-L)^{\beta }-(1-U)^{\beta }\right] u\right] ^{\frac{1%
}{\beta }}$.

\item Draw $v\sim \text{Uniform}(0,1)$; if $v\leq \left( \frac{\theta ^{\ast
}}{U}\right) ^{\alpha -1}$, set $\theta =\theta ^{\ast }$, otherwise go to
step 1.
\end{enumerate}

\begin{itemize}
\item \textbf{Algorithm: }$\alpha >1$\textbf{, }$\beta >1$\textbf{, and }$%
\frac{\alpha }{\alpha +\beta }<U$\textbf{.} If $L<\frac{\alpha }{\alpha
+\beta }$, then generate $\theta $ from $\text{B}(\alpha ,\beta )$, an
accept if $L\leq \theta \leq U$. Otherwise, define $\lambda =\frac{(\beta
-1)L-(\alpha -1)(1-L)}{L(1-L)}$. The following returns a draw from $\text{%
Beta}(\alpha ,\beta )$ truncated to $(L,U)$:
\end{itemize}

\begin{enumerate}
\item Draw $u\sim \text{Uniform}(0,1)$, and set $\theta ^{\ast }=L-\frac{1}{%
\lambda }\ln \left[ 1-\left( 1-e^{-\lambda (U-L)}\right) u\right] $.

\item Draw $v\sim \text{Uniform}(0,1)$; if $v\leq \left( \frac{\theta ^{\ast
}}{L}\right) ^{\alpha -1}\left( \frac{1-\theta ^{\ast }}{1-L}\right) ^{\beta
-1}e^{\lambda (\theta ^{\ast }-L)}$, set $\theta =\theta ^{\ast }$,
otherwise go to 1.
\end{enumerate}

This rejection algorithm for $\theta $ uses proposal density $f_{Z}(z)=\frac{%
\lambda e^{-\lambda (\theta -L)}}{1-e^{-\lambda (U-L)}}1_{(L,U)}(\theta )$.

\end{document}